\documentclass[twocolumn,showpacs,amsmath,amssymb,superscriptaddress]{revtex4-1}
\usepackage{dcolumn}
\usepackage{color}
\usepackage{graphicx}
\usepackage{amsmath}
\usepackage{bm}
\usepackage{url}

\begin{document}

\title{Structural evolution in germanium  and selenium nuclei within the 
mapped interacting boson model based on the Gogny energy density functional}

\author{K.~Nomura}
\affiliation{Physics Department, Faculty of Science, University of
Zagreb, HR-10000 Zagreb, Croatia}
\affiliation{Center for Computational
Sciences, University of Tsukuba, Tsukuba 305-8577, Japan}

\author{R.~Rodr\'iguez-Guzm\'an}
\affiliation{Physics Department, Kuwait University, 13060 Kuwait, Kuwait}

\author{L.~M.~Robledo}
\affiliation{Departamento de F\'\i sica Te\'orica, Universidad
Aut\'onoma de Madrid, E-28049 Madrid, Spain}

\date{\today}

\begin{abstract}

The shape transitions and shape coexistence in the Ge and Se isotopes 
are studied within the interacting boson model (IBM) with the 
microscopic input from the self-consistent mean-field calculation based 
on the Gogny-D1M energy density functional. The mean-field energy 
surface as a function of the quadrupole shape variables $\beta$ and 
$\gamma$, obtained from the constrained Hartree-Fock-Bogoliubov method, 
is mapped onto the expectation value of the IBM Hamiltonian with 
configuration mixing in the boson condensate state. The resultant 
Hamiltonian is used to compute excitation energies and electromagnetic 
properties of the selected nuclei $^{66-94}$Ge and $^{68-96}$Se. Our 
calculation suggests that many nuclei exhibit $\gamma$ softness. 
Coexistence between prolate and oblate, as well as between spherical 
and $\gamma$-soft, shapes is also observed. The method provides a 
reasonable description of the observed systematics of the excitation 
energy of the low-lying energy levels and transition strengths for 
nuclei below the neutron shell closure $N=50$, and provides predictions 
on the spectroscopy of neutron-rich Ge and Se isotopes with $52\leq 
N\leq 62$, where data are scarce or not available. 

\end{abstract}


\maketitle


\section{Introduction}


The evolution of the nuclear shapes has attracted considerable 
interest in nuclear structure studies \cite{BM,CasBook,cejnar2010}. 
In particular, the precise 
description of the structural evolution along different isotopic and/or isotonic chains 
as well as the associated shell effects require an accurate modeling of the 
nuclear many-body problem. Within this context, the germanium and selenium nuclei
belong to one of the most challenging regions of the nuclear chart. Their 
structure and decay patterns have been extensively studied in recent years both 
experimentally \cite{gurdal2013,corsi2013,toh2013,sun2014} and theoretically
\cite{yoshinaga2008,honma2009,kaneko2015,gaudefroy2009,niksic2014,wang2015,sarriguren2015,padilla2006,barea2009}.
Among the theoretical approximations used to study those nuclei are the 
shell model (SM) \cite{yoshinaga2008,honma2009,kaneko2015}, the energy density functional 
(EDF) framework \cite{gaudefroy2009,niksic2014,wang2015,sarriguren2015}  and the 
algebraic approach \cite{padilla2006,barea2009}. The shape transitions in the neighborhood 
of the  neutron sub-shell closure $N=40$  have also received considerable attention 
\cite{toh2013,sun2014,niksic2014,wang2015}. Moreover, the Ge and Se nuclei have been shown 
to exhibit a pronounced competition between different configurations associated 
with a variety of intrinsic shapes, i.e., shape coexistence \cite{heyde2011}. The corresponding 
spectra display low-lying excited $0^+$ energy levels which could be linked to proton intruder 
excitations across the $Z=28$ shell gap.

The EDF framework is among the most popular tools employed in microscopic nuclear structure studies. It 
allows a description of the properties of the bulk nuclear matter and the ground states of finite nuclei  
all over the nuclear chart \cite{bender2003}. Calculations are usually carried out in terms of 
the nonrelativistic Skyrme \cite{Skyrme,bender2003} and Gogny \cite{Gogny} EDFs but also  within 
the relativistic mean-field (RMF) approximation \cite{vretenar2005,niksic2011}. On the one hand, the mean-field 
approximation has already been successfully applied to nuclei with mass number
$A\approx 70-100$ \cite{gaudefroy2009,mei2012,trodriguez2014,sun2014,niksic2014,wang2015,sarriguren2015}.
On the other hand, the quantitative analysis of the collective excitations in those systems requires 
the inclusion of correlations not explicitly taken into account within the mean-field picture. Those 
correlations stem from the restoration of the symmetries (spontaneously) broken  at the mean-field level 
and/or the fluctuations in the collective coordinates. They are usually 
taken into account within the symmetry-projected generator coordinate method (GCM) 
\cite{RS,rayner2002,bender2003,niksic2011}. The symmetry projected GCM offers a reasonable starting point 
to account for the dynamical interplay between the single-particle and collective degrees of freedom 
in atomic nuclei. However, the calculations  are highly demanding from a computational
point of view, especially
in those cases where several collective coordinates should be included in the GCM ansatz. Therefore, an 
expansion in the nonlocality of the norm and Hamiltonian kernels is used to build a collective Hamiltonian 
approach  \cite{li2016} that alleviates the computational burden. At this
point the FED EXCITED VAMPIR approach of the Tubingen group used to describe
shape coexistence in some Ge and Se isotopes in 
\cite{Petrovici1988317,Petrovici1989277,Petrovici1990108,PETROVICI1992352,Petrovici2002246} has to be mentioned.

In this study, we have resorted to the fermion-to-boson mapping
procedure introduced in Ref.~\cite{nomura2008} as an alternative
approach to describe the considered Ge and Se nuclei. The method maps
the (fermionic) energy surfaces obtained with constrained mean-field
calculations onto the bosonic ones computed as the expectation value of
the interacting boson model (IBM) \cite{IBM} Hamiltonian in the boson
coherent state. 
By the mapping procedure, the parameters of the IBM Hamiltonian for each
individual nucleus are completely determined, i.e., no
phenomenological adjustment of the parameters to the experimental data
is required. 
The IBM Hamiltonian is then diagonalized and the resulting  wave
functions are used to compute the spectroscopic properties of
$^{66-94}$Ge and $^{68-96}$Se. The fermion-to-boson mapping procedure
has allowed an accurate, computationally economic and systematic
description of the shape coexistence \cite{nomura2016}, the structural
evolution in $A\approx 100$ nuclei \cite{nomura2016zr}, the quadrupole
and octupole transitions in the light actinide and rare-earth regions
\cite{nomura2014,nomura2015} as well as  odd-mass nuclei
\cite{nomura2016odd}. In this work, we demonstrate the ability of the
mapping scheme to account for the properties of the nuclei on the
neutron-deficient  side ($N\leq 50$), where there are enough
experimental data to compare with. So far, the IBM has been used in
phenomenological studies of Ge and Se nuclei
\cite{duval1983,kaup1983,padilla2006}. However, one of the advantages of
our approach is that it is able to provide predictions for unexplored
regions. We then extrapolate the method to neutron-rich nuclei with
$N=52-62$ for which, experimental data are not available. The
microscopic input is provided  by constrained Hartree-Fock-Bogoliubov
(HFB) calculations based on the finite range and density-dependent
Gogny-EDF \cite{Gogny}. In particular, we have employed the
parametrization D1M \cite{D1M}. Previous studies have shown that the
parametrization D1M essentially keeps the same predictive power as the
well tested Gogny-D1S \cite{D1S} EDF to describe a wealth of low-energy 
nuclear structure phenomena.

The paper is organized as follows. The theoretical framework used in our calculations is  outlined 
in Sec.~\ref{sec:model}. The mean-field and mapped energy surfaces are discussed in Sec.~\ref{sec:pes} while 
the  derived IBM parameters are presented in Sec.~\ref{sec:para}. We then discuss in Sec.~\ref{sec:spec} the evolution of the low-lying levels in the considered nuclei, as well as the systematics of the $B(E2)$ transition rates, spectroscopic quadrupole moments and monopole transition rates. 
We also discuss the individual level schemes for the $N=38$, 40, 42 and 60 isotones, which are representative cases of the $\gamma$ softness and/or shape coexistence.  
In  Sec.~\ref{sec:d1md1s}, we address the sensitivity 
of our predictions with respect to the particular version of the Gogny-EDF employed in the calculations. 
Finally, Sec.~\ref{sec:summary} is devoted to the conclusions and work perspectives.


\section{Description of the model\label{sec:model}}




\subsection{Self-consistent mean-field calculations}


As a first step, we have performed (constrained) HFB calculations based 
on the Gogny EDF. They provide the deformation energy surfaces for the 
considered Ge and Se nuclei as functions of the corresponding 
quadrupole deformation parameters. We have used constrains on the 
multipole operators $\hat Q_{20}$ and $\hat Q_{22}$  
\cite{robledo2008,rayner2010pt}, which are associated with the  
deformation parameters $\beta$ and $\gamma$ \cite{BM} in such a way 
that $\beta=\sqrt{4\pi/5}Q/\langle r^2\rangle$ and 
$\gamma=\tan^{-1}{Q_{22}/Q_{20}}$. Note that 
$Q=\sqrt{Q_{20}^2+Q_{22}^2}$ is the intrinsic quadrupole moment while 
$\langle r^2\rangle$ represents the mean-square radius obtained from 
the HFB state. For a more detailed account, the reader is referred to 
Ref.~\cite{rayner2010pt}. In what follows we will refer to the set of 
HFB energies, as functions of the deformation parameters  $\beta$ and 
$\gamma$, as the (mean-field) energy surface.

\subsection{The IBM Hamiltonian}
\label{IBM-Hamilt-Rayner}

To describe the spectroscopic properties of the studied nuclei, we have 
resorted to the fermion-to-boson mapping procedure introduced in 
Ref.~\cite{nomura2008}. Within such a scheme, the (fermionic) energy 
surface obtained at the Gogny-HFB level for a given nucleus is mapped 
onto the expectation value of the IBM Hamiltonian in the boson coherent 
state \cite{GK}. The parameters of the IBM Hamiltonian are then 
determined by this procedure and the excitation energies as well as the 
IBM wave functions are determined via the diagonalization of the mapped 
Hamiltonian. The transition rates are computed using such IBM wave 
functions. 

Our  IBM model comprises the collective nucleon pairs in the valence 
space with spin and parity $J^{\pi}=0^+$ (monopole $S$ pair) and $2^+$ 
(quadrupole $D$ pair). They are associated with the $J^{\pi}=0^+$ ($s$) 
and $2^+$ ($d$) bosons, respectively \cite{OAI}. The total number of 
bosons, denoted by $N_B$ amounts to  half the number of valence 
nucleons. In this study, the IBM configuration space comprises the 
proton $Z=28-50$ major shell as well as the two neutron major shells 
$N=28-50$ and $N=50-82$. Therefore, $2\le N_B\le 7$ ($3\le N_B\le 8$) 
and $3\le N_B\le 8$ ($4\le N_B\le 9$) for $^{66-82}$Ge ($^{84-94}$Ge) 
and $^{68-84}$Se ($^{86-96}$Se), respectively. In this study, for the 
sake of simplicity, no distinction has been made between the proton 
and neutron degrees of freedom. 


As will be shown later, the Gogny-HFB energy surfaces, for many of the 
considered nuclei, exhibit two minima close in energy. Within the 
mean-field picture, such minima can be  associated with the normal 
$0p-0h$ and intruder $2p-2h$ excitations across the shell gap. In the 
present case, we assume that the intruder configuration corresponds to 
the proton $2p-2h$ excitation across the shell closure $Z=28$. To 
account for the intruder configuration, the boson model space has to be 
extended. Duval and Barrett \cite{duval81} proposed a method that 
incorporates the intruder configurations by introducing several 
independent IBM Hamiltonians. As particles and holes are usually not 
distinguished, the $2p-2h$ excitation increases the boson number by 
two. The different shell-model-like spaces of $2np-2nh$ ($n=0,1$) 
configurations can be then associated with the corresponding boson 
spaces comprising $N_B+2n$ bosons. The different boson configuration 
spaces are allowed to mix via certain mixing interaction.

The Hilbert space of the configuration mixing IBM model is then defined 
as the direct sum of each unperturbed Hamiltonian, i.e., $[N_B]\oplus 
[N_{B}+2]$, where $[N_B+2n]$ denotes the unperturbed space 
corresponding to the $2np-2nh$ configurations comprising  $N_B+2n$ 
bosons. In what follows, we will simply denote the configuration 
$[N_B+2n]$ ($n=0,1$) as $[n]$. Our criterion to include the 
configuration mixing for a given nucleus is that the second-lowest 
minimum in the mean-field energy surface is clear enough so as to 
constrain the corresponding unperturbed Hamiltonian for the intruder 
configuration. According to this criterion the 
configuration-mixing has been taken into account for the nuclei 
$^{66,70-74,90-94}$Ge and $^{68-76, 90-96}$Se in this paper.

We have resorted to the configuration-mixing IBM Hamiltonian \cite{duval81}
\begin{eqnarray} 
\label{eq:ham}
 \hat H = \hat H_{0} + (\hat H_{1} + \Delta) + \hat H_{\rm mix}, 
\end{eqnarray}
where $\hat H_n$ ($n=0, 1$) is the Hamiltonian for the unperturbed 
configuration $[n]$ while $\hat H_{\rm mix}$ stands for the interaction 
mixing both spaces. In Eq.(\ref{eq:ham}), $\Delta$ represents the 
energy needed to excite one boson from one major shell to the next.

For each configuration space, we have employed the simplest form of the 
IBM-1 Hamiltonian that still simulates the essential ingredients of the 
low-energy quadrupole dynamics, i.e., 
\begin{eqnarray}
\label{eq:ham-sg}
 \hat H_n = \epsilon_n\hat n_d+\kappa_n\hat Q\cdot\hat Q + \kappa^{\prime}_n\hat V_{ddd}. 
\end{eqnarray}
The first term $\hat n_d = d^{\dagger}\cdot\tilde d$ in 
Eq.(\ref{eq:ham-sg}), is the $d$-boson number operator and $\epsilon_n$ 
is the single $d$-boson energy in the $[n]$ space. The second term 
represents the quadrupole-quadrupole interaction with strength 
parameter $\kappa_n$. The quadrupole operator $\hat Q$ in boson space reads $\hat 
Q=s^{\dagger}\tilde d + d^{\dagger}s + \chi_n[d^{\dagger}\times\tilde 
d]^{(2)}$, where $\chi_n$ is a parameter.  The third term stands for a 
specific three-body interaction among $d$ bosons, with strength  
$\kappa^{\prime}_n$, which is required to describe $\gamma$-soft 
systems \cite{nomura2012tri}. It takes the form
\begin{eqnarray}
 \hat V_{ddd} = [[d^{\dagger}\times d^{\dagger}\times d^{\dagger}]^{(L)}\times [[\tilde d\times\tilde d\times\tilde d]^{(L)}]^{(0)}, 
\end{eqnarray}
where the symbol $\times$ represents a tensor coupling and $L$ is the total 
angular momentum of the boson system. In our calculations, we have only 
included  the term with $L=3$ as it gives rise to a stable minimum at 
$\gamma\approx 30^{\circ}$. The mixing interaction term $\hat H_{\rm 
mix}$ reads
\begin{eqnarray}
 \hat H_{\rm mix}=\omega_s s^{\dagger}s^{\dagger} + \omega_d d^{\dagger}\cdot d^{\dagger} + (h.c.), 
\end{eqnarray}
where $\omega_s$ and $\omega_d$ are strength parameters. For simplicity, 
we have assumed  $\omega_s=\omega_d\equiv\omega$.

To associate a Gogny-HFB energy surface with the corresponding 
configuration-mixing IBM Hamiltonian Eq.~(\ref{eq:ham}), an extended 
boson coherent state 
\begin{eqnarray}
 |\Phi(\beta,\gamma)\rangle=|\Phi(N_0,\beta,\gamma)\rangle\oplus |\Phi(N_1,\beta,\gamma)\rangle, 
\end{eqnarray}
has been introduced with $N_{n}=N_{B}+2n$ ($n=0,1$). For each unperturbed 
configuration space $|\Phi(N_{n},\beta,\gamma)\rangle$ ($n=0,1$), the 
coherent state is taken in the form
\begin{eqnarray}
|\Phi(N_{n},\beta,\gamma)\rangle=\frac{1}{\sqrt{N_{n}!}}(\lambda^{\dagger})^{N_{n}}|0\rangle
\end{eqnarray}
where $|0\rangle$ denotes the inert core and 
\begin{eqnarray}
 \lambda^{\dagger}=s^{\dagger}+\beta_B\cos{\gamma_B}d^{\dagger}_0+\frac{1}{\sqrt{2}}\beta_B\sin{\gamma_B}(d^{\dagger}_{+2}+d^{\dagger}_{-2}). 
\end{eqnarray}
On the other hand, $\beta_B$ and $\gamma_B$ are the boson analogs of the 
quadrupole deformation parameters $\beta$ and $\gamma$ within the 
geometrical collective model  \cite{BM}.

The expectation value of the total Hamiltonian $\hat H$ in the coherent 
state $|\Phi(\beta,\gamma)\rangle$ leads to a $2\times 2$ matrix \cite{frank04}: 
\begin{eqnarray}
\label{eq:pes}
  {\cal E}=\left(
\begin{array}{cc}
E_{0}(\beta,\gamma) & \Omega(\beta) \\
\Omega(\beta) & E_{1}(\beta,\gamma)+\Delta \\
\end{array}
\right), 
\end{eqnarray}
with diagonal and off-diagonal elements accounting for the expectation 
values of the unperturbed and mixing terms, respectively. The two 
eigenvalues of ${\cal E}$ correspond to specific energy surfaces. It is 
customary to take the lower-energy one \cite{frank04} as the IBM $(\beta,\gamma)$-energy.

The diagonal matrix element $E_n(\beta,\gamma)$ is given by
\begin{eqnarray}
\label{eq:pes-detail1}
E_{n}(\beta,\gamma)&=&\frac{k_1+k_2\beta_{n}^2}{1+\beta_{n}^2}
+\frac{k_3\beta_{n}^2+k_4\beta_{n}^3\cos{3\gamma}  
+ k_5\beta_{n}^4}{(1+\beta_{n}^2)^2} \nonumber \\
& + & \frac{k_6\beta_{n}^6\sin^2{3\gamma}}{(1+\beta_{n}^2)^3}
\end{eqnarray}
where $k_1=5\kappa_n N_{n}$, $k_2=[\epsilon_n+\kappa_n(1+\chi^2_{n})]N_{n}$, 
$k_3=4\kappa_n N_{n}(N_{n}-1)$, $k_4=-4\kappa_n\sqrt{2/7}N_{n}(N_{n}-1)\chi_{n}$, 
$k_5=(2/7)\kappa_n N_{n}(N_{n}-1)\chi_{n}^2$ and 
$k_6=(1/30)\kappa^{\prime}_{n}N_{n}(N_{n}-1)(N_{n}-2)$. 
Moreover, the non-diagonal matrix element reads
\begin{eqnarray}
\label{eq:pes-detail2}
\Omega(\beta)=\omega\sqrt{(N_{B}+1)(N_{B}+2)}
\Big[\frac{1+\beta_{0}\beta_{1}}{\sqrt{(1+\beta_{0}^2)(1+\beta_{1}^2)}}\Big]^{N_{B}}. 
\end{eqnarray}
Note that, in Eqs.~(\ref{eq:pes-detail1}) and (\ref{eq:pes-detail2}),  
$\beta_{n}$ represents the bosonic deformation parameter  for each 
unperturbed space $[n]$. It is related to the Gogny-HFB one as 
$\beta_{n}=C_{n}\beta$. The constant  $C_n$ is  also determined  by 
fitting  the (fermionic) Gogny-HFB energy surface to the (bosonic) IBM 
one. To this end, one requires  that the position of the minimum, for 
each unperturbed configuration, is reproduced. Both 
Eqs.~(\ref{eq:pes-detail1}) and (\ref{eq:pes-detail2}) are similar to 
the ones employed in our previous studies 
\cite{nomura2012sc,nomura2013hg,nomura2016zr} within the IBM-2 framework. 


\subsection{Derivation of the IBM parameters: the fitting procedure}


The  Hamiltonian in Eq.~(\ref{eq:ham}) contains 10 parameters. They have 
been determined along the following lines:
 
\begin{description}
 \item[Step 1] Each unperturbed Hamiltonian is determined by using
	    the procedure of Refs.~\cite{nomura2008,nomura2010,nomura2016zr}. Here,  
	    each diagonal matrix element $E_{n}$ in Eq.~(\ref{eq:pes})
	    is fitted to the corresponding mean-field minimum. 
	    The
	    normal $[n=0]$ configuration is assigned to the mean-field minimum
	    with the smallest deformation while the 
	    $[n=1]$ configuration is assigned to
	    the HFB minimum with the larger deformation. In this way, each 
	    unperturbed Hamiltonian is determined
	    independently.		       	   	    
\item[Step 2] The energy offset $\Delta$ is determined so that the energy difference 
              between the two minima (Step 1) of the  Gogny-HFB energy surface is
	          reproduced. 	    
\item[Step 3] Finally, the strength parameter $\omega$ of the mixing interaction 
              term $\hat H_{\rm mix}$  is determined so as to reproduce the
	          shapes of the barriers between the minima \cite{nomura2012sc,nomura2013hg}.
\end{description}

In Step 1, note that the link of the $0p-0h$ and $2p-2h$
configurations with the small and large deformation minima, respectively, is
based on the assumption that the well-established interpretation of 
shape coexistence in the neutron-deficient lead region
\cite{bengtsson1987,bengtsson1989,nazarewicz1993} also holds here. In these 
references, the $0^+_1$ ground state is associated with a weakly-deformed oblate shape 
and the intruder $0^+_2$ state with a prolate shape with larger
deformation. 

Once the IBM parameters for each of the considered nuclei are determined, the Hamiltonian $\hat H$
is diagonalized in the $[0]\oplus [1]$ space by using the code IBM-1 \cite{IBM1}. 
The IBM wave functions resulting from the diagonalization are then used to compute electromagnetic 
properties that could be considered as signatures of shape coexistence and/or shape transitions 
such as, the $B(E2)$ transition probabilities, the spectroscopic quadrupole moments $Q_{sp}$ and  
the $\rho^2({E0})$ values between $0^+$ states. The $B(E2)$ transition probabilities read

\begin{eqnarray}
 B(E2; J_i\rightarrow J_f)=\frac{1}{2J_i+1}|\langle J_f||\hat T^{(E2)}||J_i\rangle|^2, 
\end{eqnarray}
where $J_i$ and $J_f$ are the spins of initial and final states, respectively. On the other hand, the 
spectroscopic quadrupole moments and the  $\rho^2({E0})$ values are computed as 
\begin{eqnarray}
 Q_{sp}=\sqrt{\frac{16\pi}{5}}
\left(
\begin{array}{ccc}
 J & 2 & J \\
 -J & 0 & J \\
\end{array}
\right)
\langle J||\hat T^{(E2)}||J\rangle, 
\end{eqnarray}
and
\begin{eqnarray}
\label{eq:rhoe0}
 \rho^2(E0; 0^+_i\rightarrow 0^+_f) = \frac{Z^2}{R_0^4}|\langle 0^+_f||\hat T^{(E0)}||0^+_i\rangle|^2
\end{eqnarray}
where $R_0=1.2\,A^{1/3}$ fm. The E0 and E2 operators take the form 
$\hat T^{(E0)}=\sum_{n=0,1}e_{0,n}\hat n_d$ and 
$\hat T^{(E2)}=\sum_{n=0,1}e_{2,n}\hat Q$, respectively. 
For the  effective charges we have assumed
$e_{0,0}=e_{0,1}\equiv e_{0}$ and $e_{2,0}=e_{2,1}\equiv e_{2}$. 
Their numerical values have been fitted 
as to reproduce the experimental 
$B(E2; 2^+_1\rightarrow 0^+_1)$ \cite{data} and 
$\rho^2(E0; 0^+_2\rightarrow 0^+_1)$ \cite{kibedi2005} 
values for the $N=42$ and $N=40$ nuclei, respectively.


\section{Energy surfaces\label{sec:pes}}



\begin{figure*}[htb!]
\begin{center}
\includegraphics[width=0.8\linewidth]{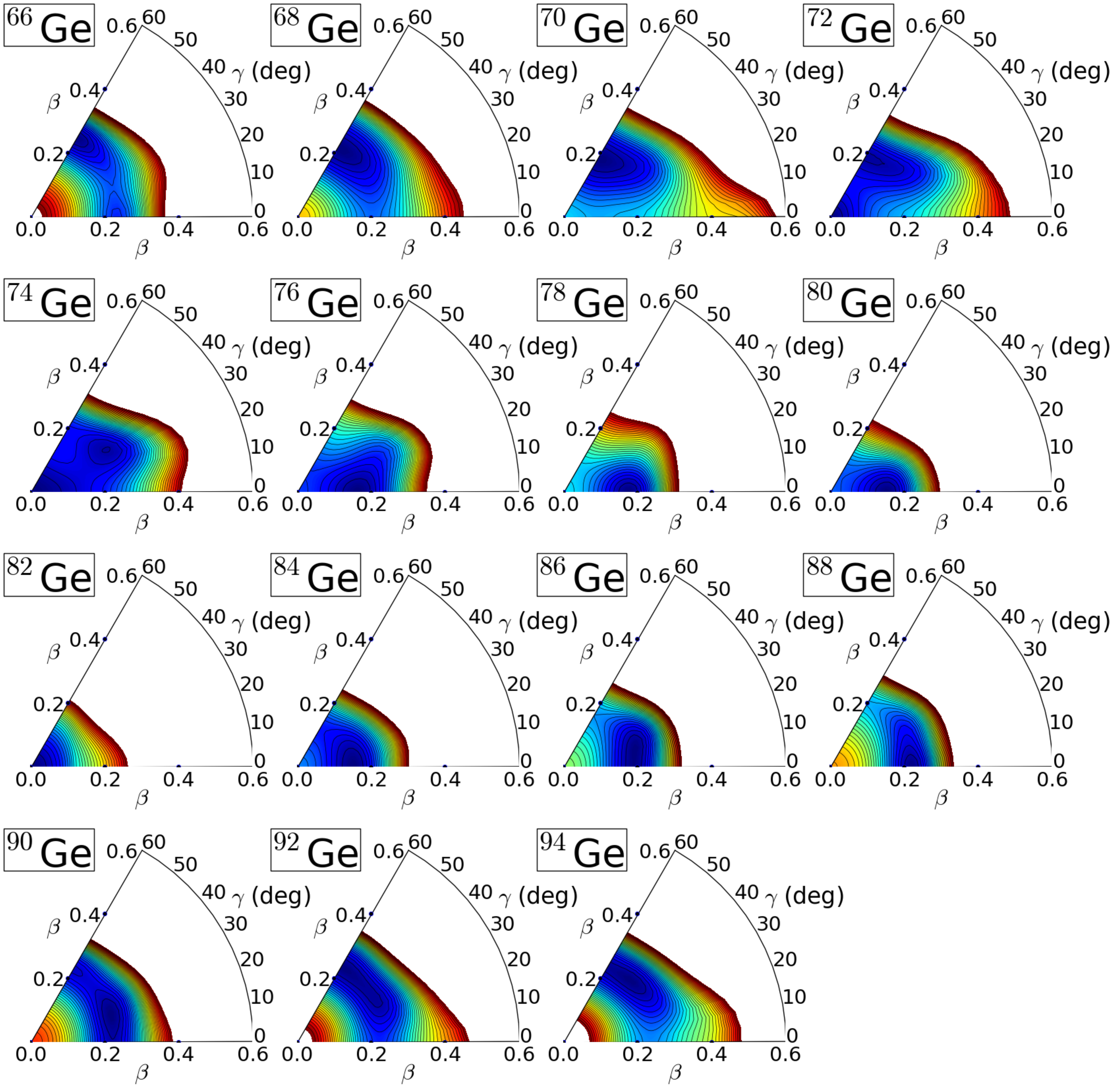}
\caption{(Color online) Mean-field energy surfaces for the nuclei $^{66-94}$Ge. 
Results have been obtained with the Gogny-D1M EDF. The energy difference 
between neighboring contours is 100 keV. }
\label{fig:pes-ge-hfb}
\end{center}
\end{figure*}


\begin{figure*}[htb!]
\begin{center}
\includegraphics[width=0.8\linewidth]{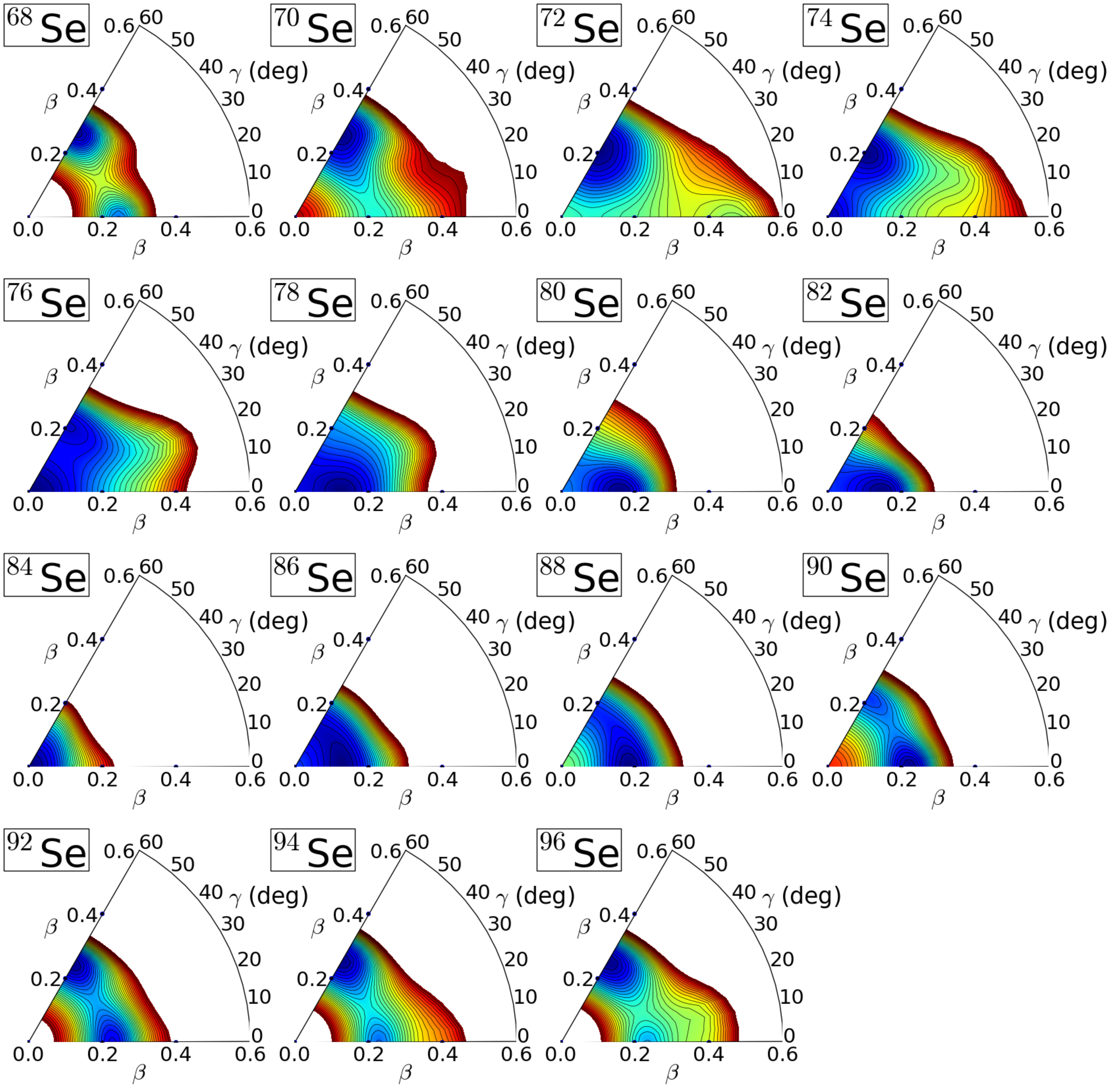}
\caption{(Color online) The same as in Fig.~\ref{fig:pes-ge-hfb}, but 
for the  nuclei $^{68-96}$Se.}
\label{fig:pes-se-hfb}
\end{center}
\end{figure*}


\subsection{Gogny-D1M energy surfaces}


The Gogny-D1M energy surfaces, obtained for the nuclei $^{66-94}$Ge and 
$^{68-96}$Se, are shown in Figs.~\ref{fig:pes-ge-hfb} and 
\ref{fig:pes-se-hfb}. Similar results have been obtained with the 
parametrization D1S of the Gogny-EDF and therefore they will not be 
discussed in detail in this section. As can be seen from 
Fig.~\ref{fig:pes-ge-hfb}, the nucleus $^{66}$Ge exhibits coexisting 
prolate and oblate minima with $\beta\approx 0.2$. The prolate minimum 
becomes less pronounced in both $^{68,70}$Ge. A shape transition is 
observed between the ground-state shapes of $^{70}$Ge and $^{72}$Ge. For
the latter, a spherical 
minimum emerges and becomes the ground state at the mean-field
level. Such a spherical ground state could be associated with the $N=40$
neutron sub-shell closure. 
Furthermore, a close-lying oblate minimum is also observed in the 
energy surface of $^{72}$Ge. In the case of $^{74}$Ge, one observes a 
coexistence between the spherical ground state and a triaxial minimum 
with $\gamma\approx 30^{\circ}$. A single prolate minimum, which is 
notably $\gamma$-soft, is found for $^{76}$Ge. For higher neutron 
numbers, the minimum moves gradually from prolate to spherical, 
reflecting the proximity of the $N=50$ neutron shell-closure. A prolate 
minimum develops from $^{82}$Ge to $^{88}$Ge and becomes 
$\gamma$-softer as a function of the neutron number. On the other hand, 
a shallow oblate minimum is found for $^{90}$Ge. An oblate and 
$\gamma$-soft ground state is predicted for the isotopes $^{92,94}$Ge. 
As can be seen from Fig.~\ref{fig:pes-se-hfb}, a similar structural 
evolution is predicted  for the studied Se nuclei. Our Gogny-D1M HFB 
trends agree well with previous results obtained within the 
relativistic mean-field (RMF) approximation \cite{niksic2014}. A 
coexistence between spherical and oblate configurations has also been 
found for $^{70, 72}$Se \cite{PhysRevLett.100.102502} and  $^{74}$Se 
within the 5D collective Hamiltonian approach based 
on the Gogny-D1S EDF \cite{gaudefroy2009}.


\begin{figure*}[htb!]
\begin{center}
\includegraphics[width=0.8\linewidth]{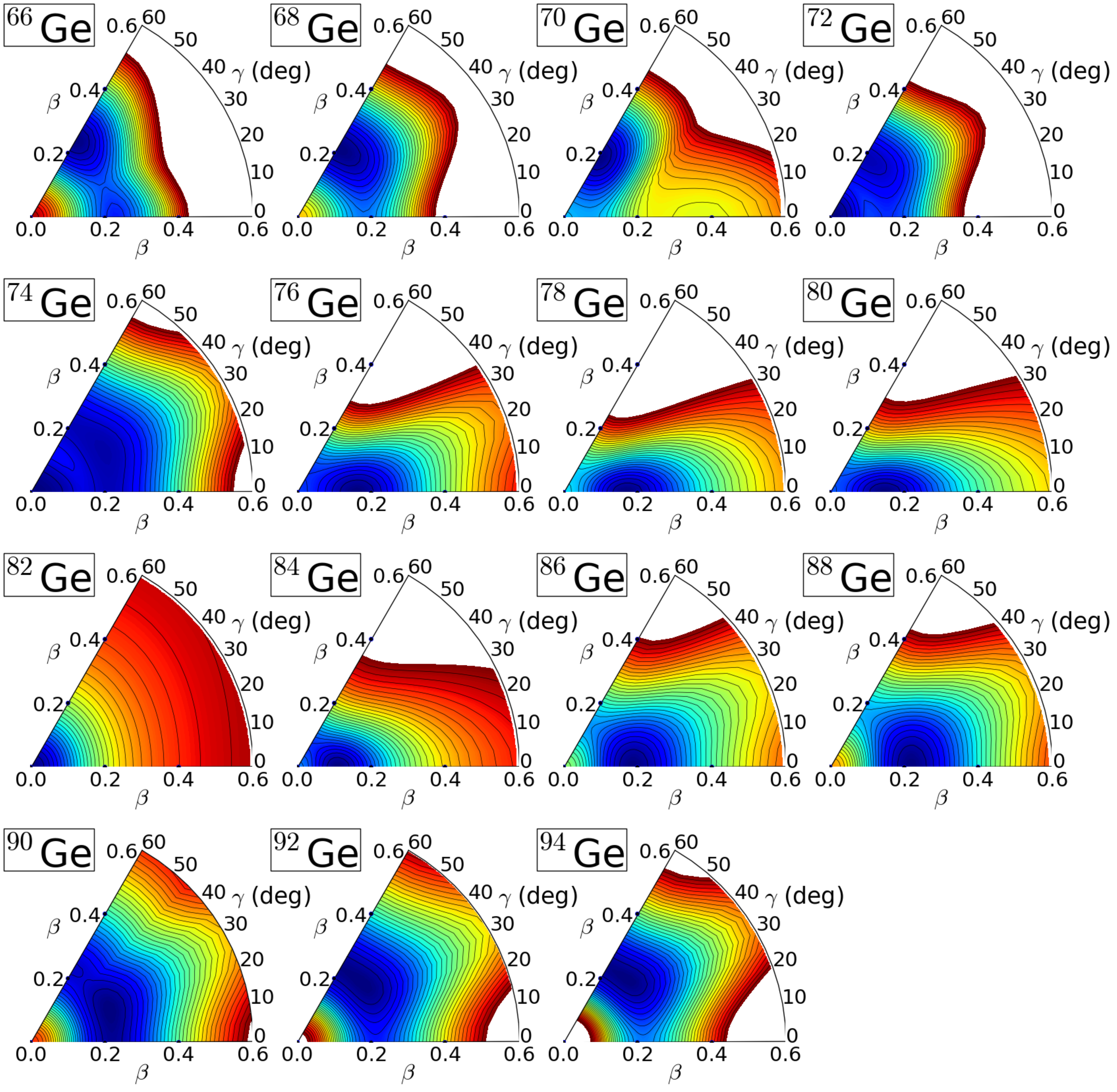}
\caption{(Color online) The same as in  Fig.~\ref{fig:pes-ge-hfb}, 
but for the mapped IBM energy surfaces.}
\label{fig:pes-ge-mapped}
\end{center}
\end{figure*}


\begin{figure*}[htb!]
\begin{center}
\includegraphics[width=0.8\linewidth]{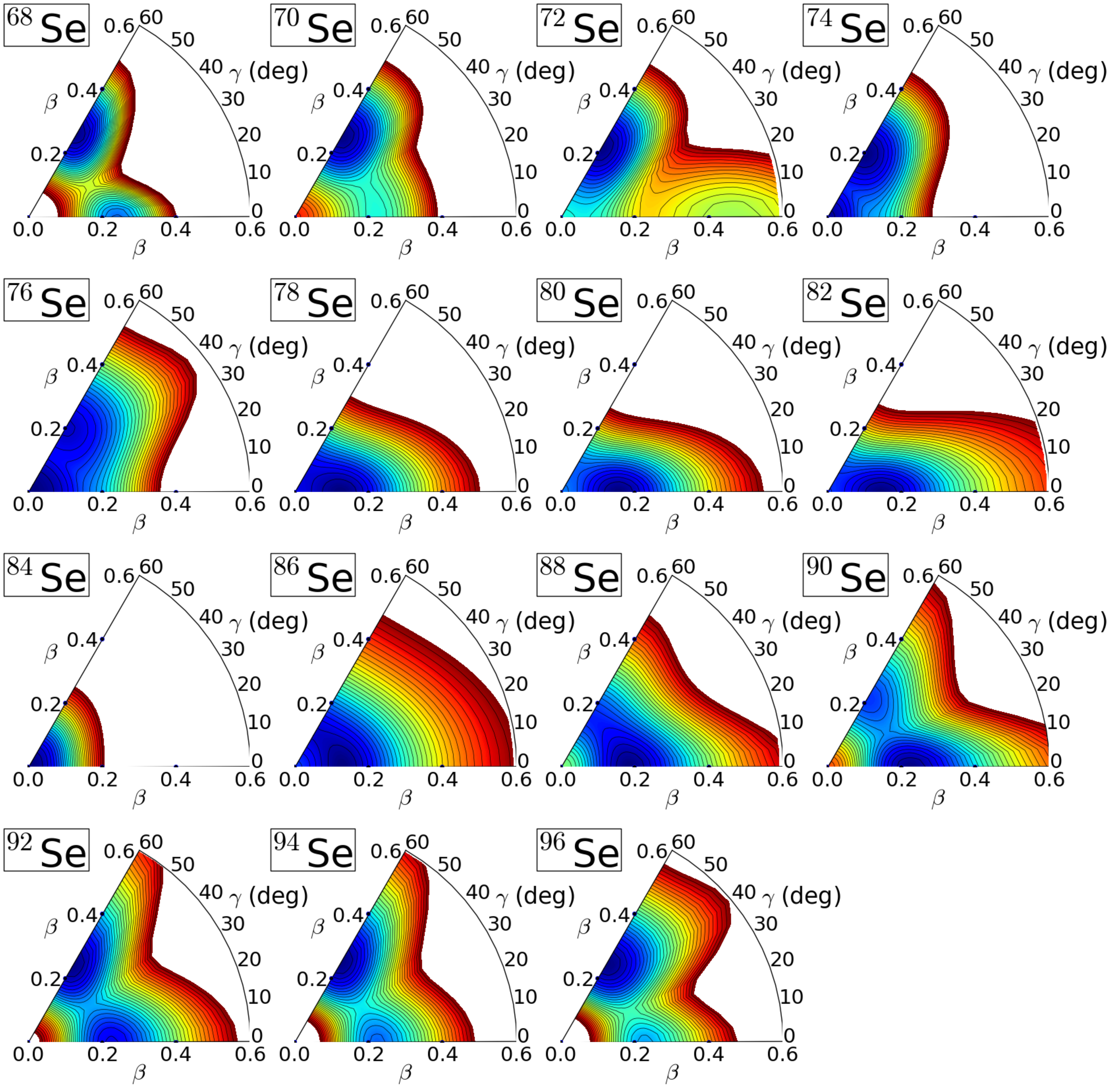}
\caption{(Color online) The same as in Fig.~\ref{fig:pes-se-hfb}, but for 
the mapped IBM energy surfaces.}
\label{fig:pes-se-mapped}
\end{center}
\end{figure*}


\subsection{Mapped IBM energy surfaces}


In Figs.~\ref{fig:pes-ge-mapped} and \ref{fig:pes-se-mapped} we have 
plotted, the IBM energy surfaces obtained by mapping the Gogny-D1M ones 
already shown in  Figs.~\ref{fig:pes-ge-hfb} and \ref{fig:pes-se-hfb}. 
First we realize that, compared with the mean-field energy surfaces, 
the IBM ones are generally more flat in those regions of the $\beta-\gamma${}
plane away from the ground 
state minimum. This behavior arises as a consequence of the limited 
number of nucleon pairs (bosons) comprising the IBM valence space but also 
because the Hamiltonian used for each configuration space 
Eq.~(\ref{eq:ham-sg}) is too simple to reproduce every detail of the 
fermionic energy surfaces \cite{nomura2008,nomura2010}. In order to 
determine the IBM Hamiltonian we have reproduced the location and depth 
of the energy minimum as well as the curvatures along both the $\beta$ 
and $\gamma$ directions around the minimum. Furthermore, we have also 
reproduced the topology of the barriers separating the different 
minima. With this in mind, ones observes from 
Figs.~\ref{fig:pes-ge-mapped} and \ref{fig:pes-se-mapped} that the
trends observed as functions of the neutron number N in the  
mapped energy surfaces mimic quite well the 
ones found in the Gogny-D1M case.


\section{Evolution of the derived IBM parameters\label{sec:para}}



\begin{figure}[htb!]
\begin{center}
\includegraphics[width=\columnwidth]{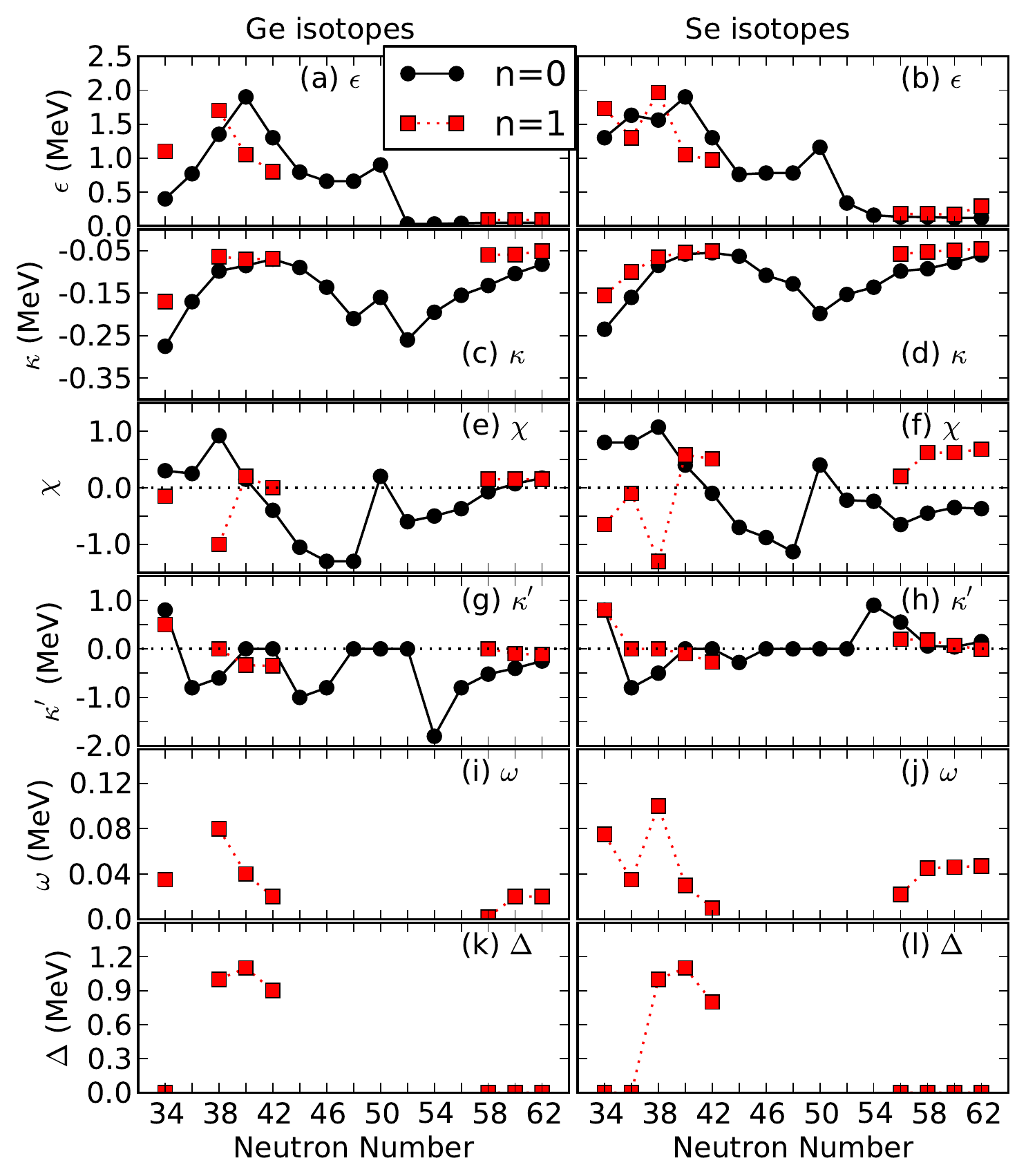}
\caption{(Color online) The IBM parameters $\epsilon$, $\kappa$, $\chi$, 
$\kappa^{\prime}$, $\omega$ and  $\Delta$ are depicted, as functions of 
the neutron number, for the $[n=0]$ and $[n=1]$ configurations. For more 
details, see the main text.}
\label{fig:para}
\end{center}
\end{figure}

In Fig.~\ref{fig:para} we have depicted the parameters of the IBM 
Hamiltonian, obtained via the fermion-to-boson mapping procedure, as 
functions of the neutron number. The decrease of the single $d$-boson 
energy $\epsilon$ [panels (a) and (b)] when moving towards the open-shell region, 
reflects  the emergence of collectivity.  For both the normal and 
intruder configurations, the parameter $\epsilon$ increases when 
approaching the neutron sub-shell $N\approx 40$ and the magic number 
$N=50$. On the other hand, the  $\epsilon$ values for neutron-rich Ge 
and Se  nuclei with $N\ge 52$  are rather small. The strength $\kappa$ 
of the  quadrupole-quadrupole interaction is shown in panels (c) and 
(d). It exhibits a gradual decrease when moving away from the shell 
closure, a trend already found in previous IBM studies 
\cite{OAI,nomura2010}. Note that around $N=40$, the strength  $\kappa$ 
is much less sensitive to the neutron number than the parameter 
$\epsilon$. The parameter $\chi$ determines whether a nucleus is 
prolate ($\chi<0$), oblate ($\chi>0$) or $\gamma$-soft ($\chi\approx 
0$). As can be seen from panels (e) and (f),  for the normal 
configuration in Ge nuclei, it changes sign from $N=38$ to 44 which is 
consistent, with the oblate-to-prolate transition observed for the 
minimum of the Gogny-D1M  and mapped energy surfaces.

The strength of the three-body boson term also reflects  
$\gamma$-softness. In particular, a negative value of $\kappa^{\prime}$ 
creates a stable triaxial minimum at $\gamma=30^{\circ}$ whereas a 
positive value leads to stiffness along the  $\gamma$ direction (see 
Eq.~(\ref{eq:pes-detail1})). From panels (g) and (h) one realizes that, 
for several of the considered Ge isotopes, the  $\kappa^{\prime}$ 
values for the normal configurations are negative and notably large in 
magnitude. This reflects that the Gogny-D1M energy surfaces are 
generally  $\gamma$-softer for Ge than  for Se nuclei. The mixing 
strength  $\omega$ [panels (i) and (j)] and the energy off-set $\Delta$ 
[panels (k)  and (l)] are of the same order of magnitude as those 
obtained in previous IBM configuration mixing calculations 
\cite{padilla2006}. Note that the $\omega$ values are particularly 
large for $N=38$ in both the Ge and Se isotopic chains. 
In this case, the two minima observed in the Gogny-D1M energy surface are
rather well separated from each other along the $\gamma$ direction
and therefore large $\omega$ values are required.


\section{Results for  spectroscopic properties\label{sec:spec}}


\subsection{Systematics of the excitation energies\label{sec:energy}}


\begin{figure}[htb!]
\begin{center}
\includegraphics[width=\columnwidth]{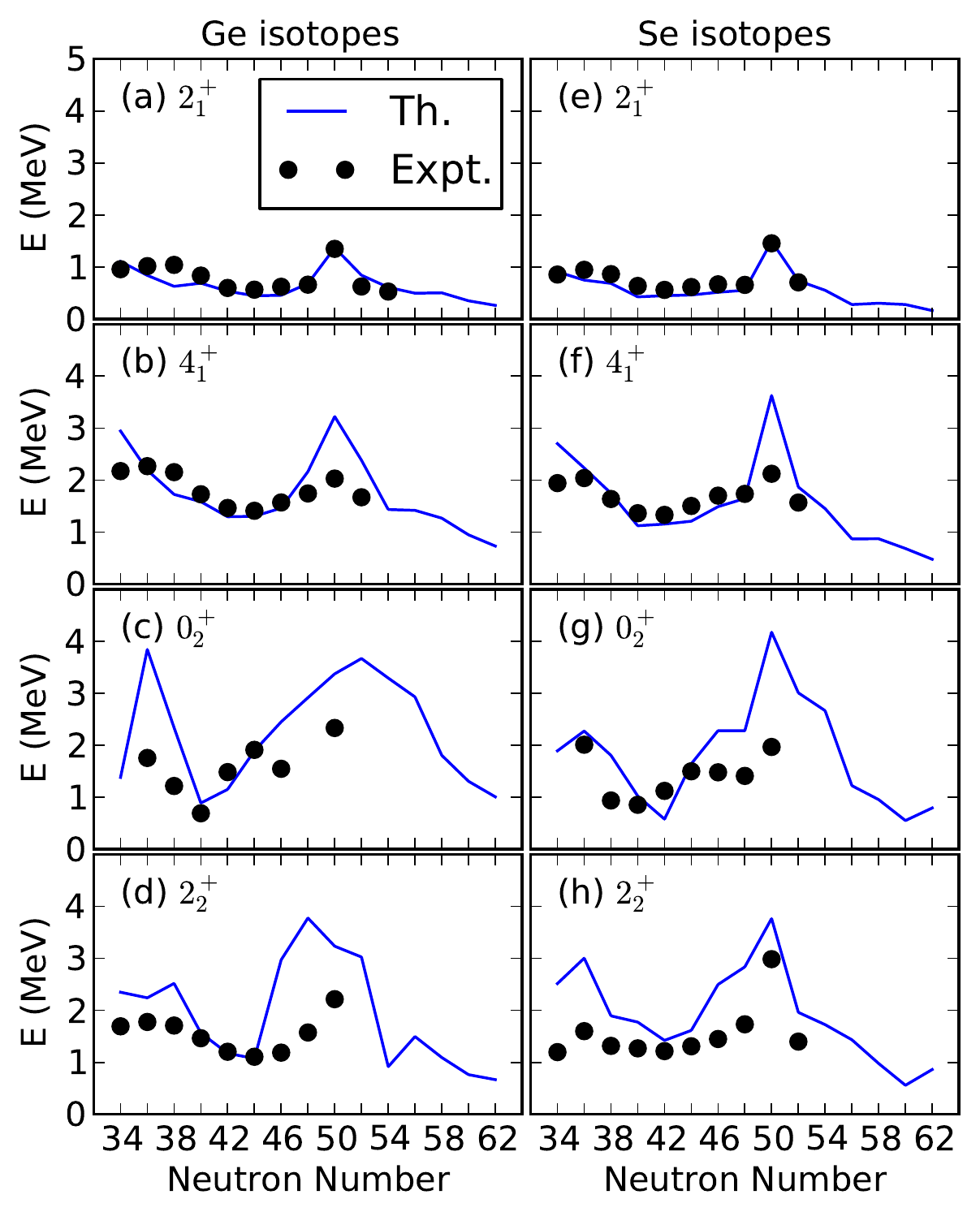}
\caption{(Color online) The $2^+_1$, $4^+_1$, $0^+_2$ and $2^+_2$ 
excitation energies obtained in the diagonalization of the
mapped IBM Hamiltonian are plotted as functions of the neutron number, for 
the Ge and Se nuclei, along with 
the available experimental data \cite{data}.}
\label{fig:energies}
\end{center}
\end{figure}

\begin{figure}[htb!]
\begin{center}
\includegraphics[width=\columnwidth]{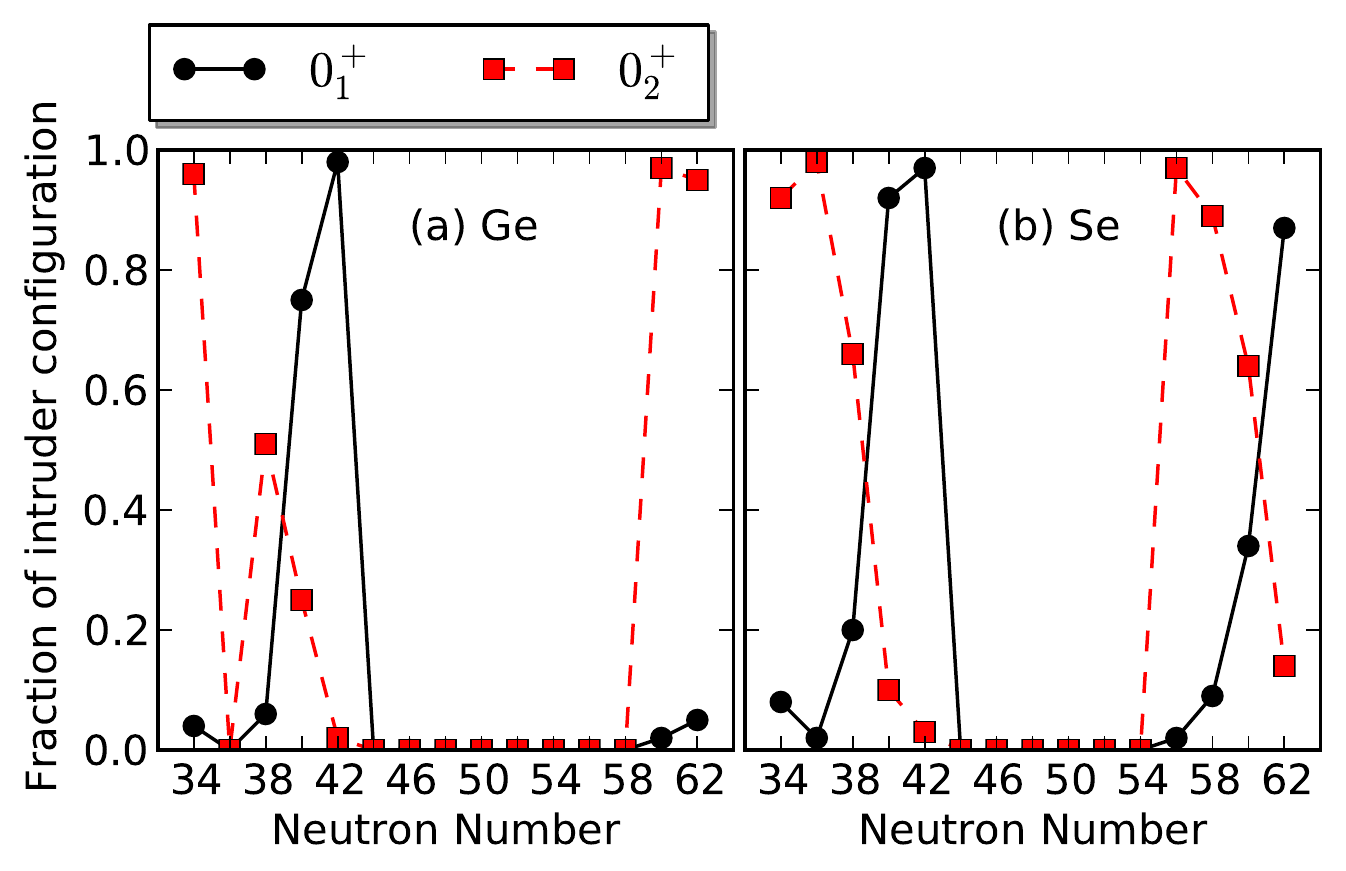}
\caption{(Color online) Fraction of the intruder configuration 
in the IBM $0^+_1$ and $0^+_2$ wave functions. For more 
details, see the main text.}
\label{fig:frac}
\end{center}
\end{figure}

The excitation energies of the $2^+_1$, $4^+_1$, $0^+_2$ and $2^+_2$ 
states obtained in this work, are displayed in Fig.~\ref{fig:energies} 
as functions of the neutron number. They are compared with the 
available experimental data \cite{data}. As can be seen, our 
calculations provide a reasonable agreement with the experimental 
systematics, especially for the yrast states. The $E(2^+_1)$ energy
[panels (a) and (e)] can be regarded 
as one of the best signatures for a shape/phase transition 
\cite{cejnar2010}. For both Ge and Se nuclei, the computed $E(2^+_1)$ 
energies decrease as one approaches $N=40$. In the case of Ge isotopes, 
this is at variance with the experiment. This discrepancy could be 
attributed to the  $N=40$ neutron sub-shell closure  not  explicitly 
taken into account  in our  calculations. Moreover, the $E(2^+_1)$ 
values exhibit a pronounced peak at $N=50$. In the case of the 
$E(4^+_1)$ excitation energies [panels (b) and (f)], our results 
overestimate the experimental ones around $N=50$. This could be linked 
to the limited IBM configuration space comprising only $s$ and $d$ 
bosons. The inclusion of the $J=4^+$ ($G$) pair in the IBM model could 
improve the agreement with the experiment but lies out of the scope of 
this study. Work along these lines is in progress and will be reported 
elsewhere.

The appearance of low-lying  $0^+_2$ states is often attributed to 
intruder excitations and regarded as a signature of shape coexistence 
\cite{heyde2011}. The predicted $E(0^+_2)$ energies are plotted in 
panels (c) and (g). They display a pronounced decrease towards 
$N\approx 40$. This correlates well with the  shape coexistence 
observed in the underlying Gogny-D1M energy surfaces around this 
neutron number. The overestimation of the $E(0^+_2)$ energy in the case 
of $^{68}$Ge is due to the fact that a configuration mixing calculation 
has not been carried out in this case. The fraction of the intruder 
configuration in the IBM $0^+_1$ and $0^+_2$ wave functions for Ge and 
Se nuclei is plotted in panels (a) and (b) of Fig.~\ref{fig:frac} as a 
function of the neutron number $N$. From the plots, one realizes that, 
for both Ge and Se, the $0^+_1$ and $0^+_2$ states at $N=38$ mainly 
arise from the normal and intruder configuration, respectively. At 
$N=40$ and 42, in contrast, the $0^+_1$ state is dominated by the 
intruder configuration, while the $0^+_2$ state is almost purely made 
of the normal configuration. Coming back to Fig.~\ref{fig:energies} 
[panels (c) and (g)], for the  considered neutron-rich nuclei, several 
examples of   low-lying $0^+_2$ states are found beyond the $N=50$ 
shell closure. Finally,  from the plots in panels (d) and (h), we 
conclude that  our calculations  lead to a reasonable description of 
the energies of the $2^+_2$ states which are, either interpreted as 
bandheads of the quasi-$\gamma$ bands or as members of the $0^+_2$ 
bands. 

We note, in both Ge and Se isotopes, that the predicted excitation energies
of the non-yrast states $E(0^+_2)$ and $E(2^+_2)$ are generally higher
than the experimental values especially for $46\leq N\leq 50$. 
This discrepancy has been commonly observed in our previous calculations
for other mass regions using the HFB-to-IBM mapping procedure
(see, e.g., Ref.~\cite{nomura2016zr}) and could be, in most cases,
attributed to the restricted model space of the IBM when the shell
closure is approached.


\subsection{Electromagnetic properties\label{sec:em}}


\subsubsection{$B(E2)$ transition rates}

\begin{figure}[htb!]
\begin{center}
\includegraphics[width=\columnwidth]{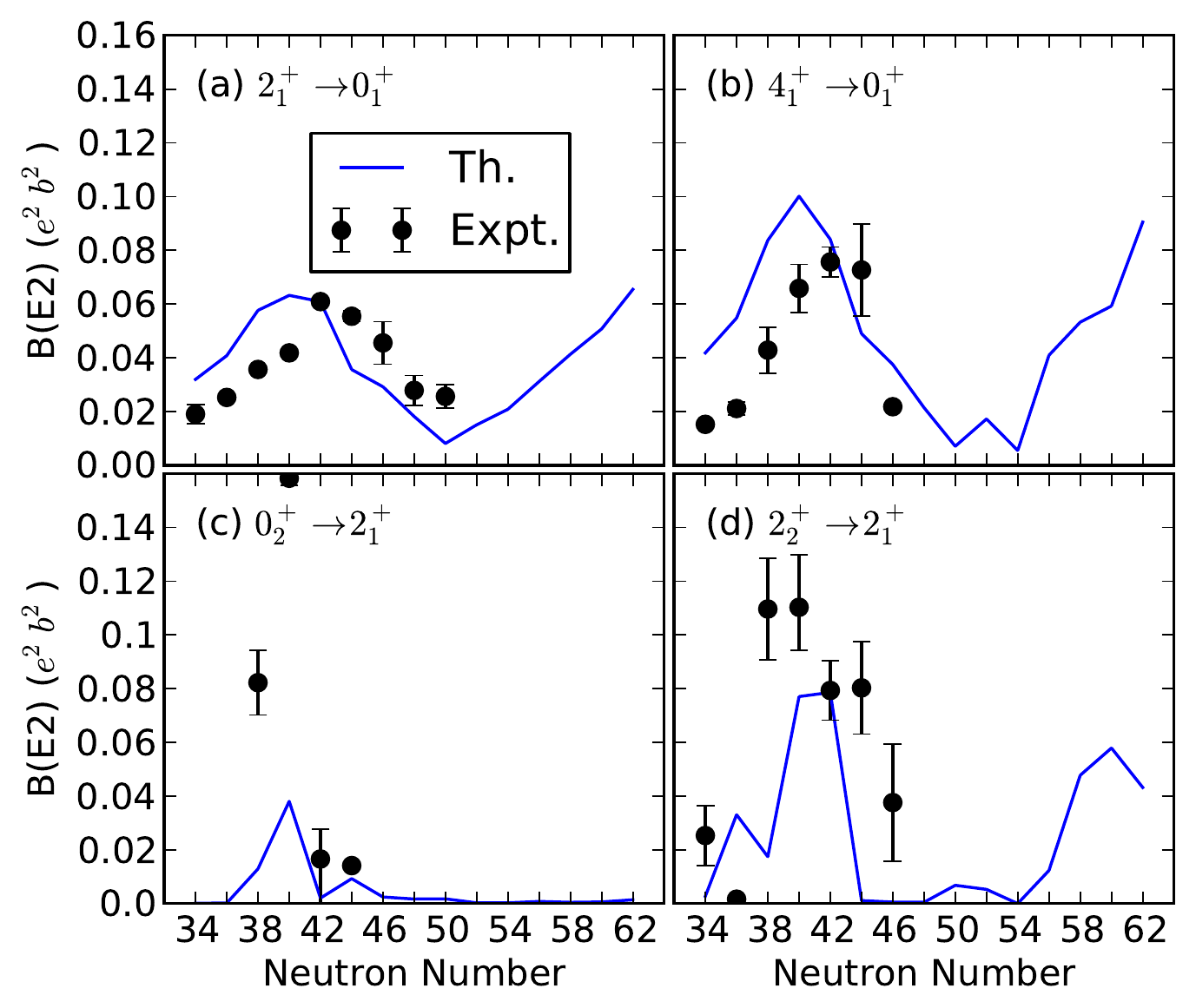}
\caption{(Color online) The $B(E2; 2^+_1\rightarrow 0^+_1)$, 
$B(E2; 4^+_1\rightarrow 2^+_1)$, $B(E2; 0^+_2\rightarrow 2^+_1)$ 
and $B(E2; 2^+_2\rightarrow 2^+_1)$ transition probabilities 
obtained for Ge isotopes are plotted as functions 
of the neutron number. Experimental data have been 
taken from Ref.~\cite{data}.}
\label{fig:ge_e2}
\end{center}
\end{figure}


\begin{figure}[htb!]
\begin{center}
\includegraphics[width=\columnwidth]{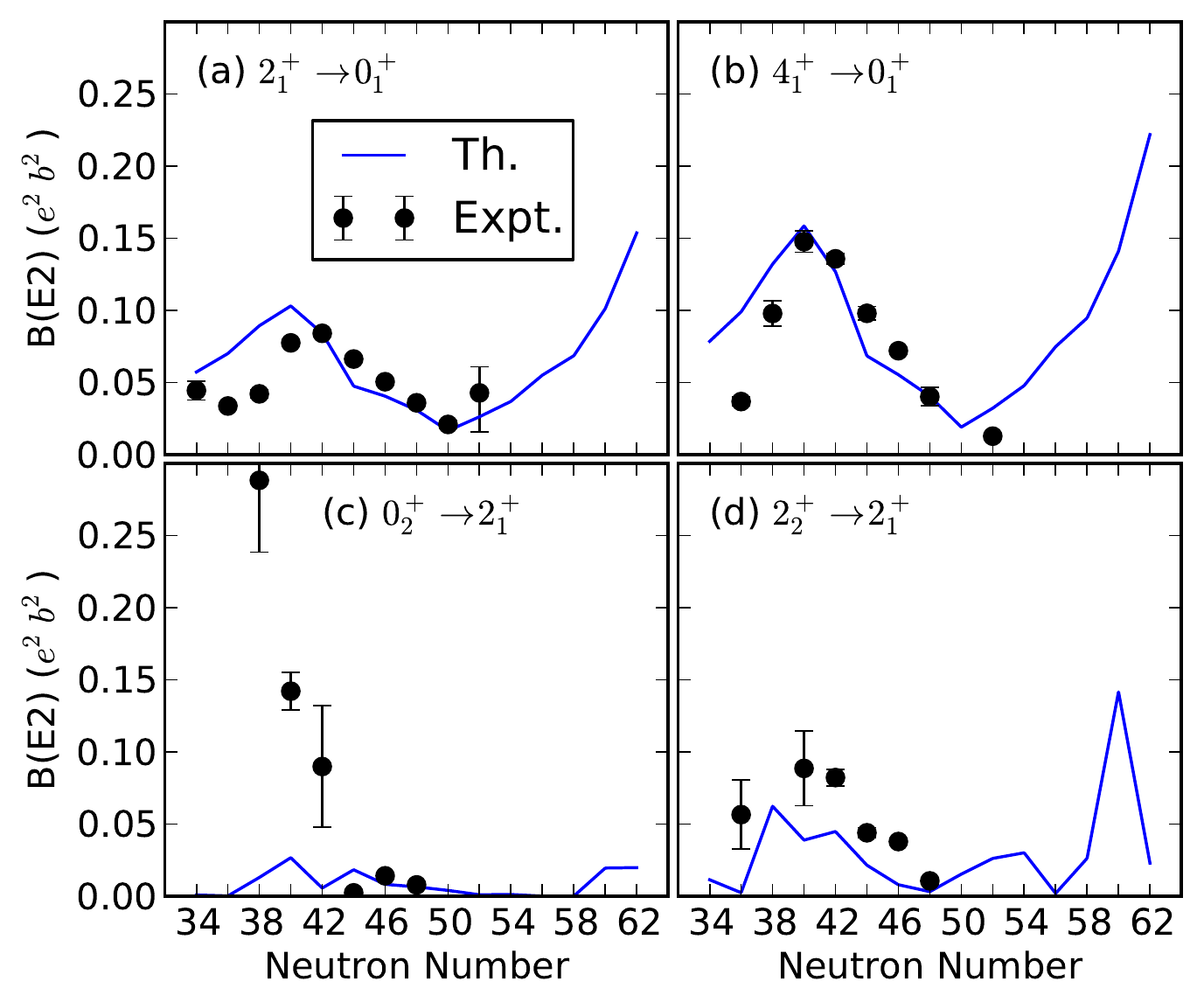}
\caption{(Color online) The same as in Fig.~\ref{fig:ge_e2}, but for 
the Se isotopes. Experimental data have been taken from Ref.~\cite{data}.}
\label{fig:se_e2}
\end{center}
\end{figure}

The transition probabilities $B(E2; 2^+_1\rightarrow 0^+_1)$, $B(E2; 
4^+_1\rightarrow 2^+_1)$, $B(E2; 0^+_2\rightarrow 2^+_1)$ and $B(E2; 
2^+_2\rightarrow 2^+_1)$ are depicted in Figs.~\ref{fig:ge_e2} and 
\ref{fig:se_e2} for Ge and Se nuclei, respectively. The maximum $B(E2; 
2^+_1\rightarrow 0^+_1)$ value is reached around $N=40$ where the 
deformation is the largest [panel (a) of Figs.~\ref{fig:ge_e2} and 
\ref{fig:se_e2}]. The agreement between our results and the experimental 
data for Ge and Se nuclei is fairly good. A similar trend is also 
found for the $B(E2; 4^+_1\rightarrow 2^+_1)$ transition rates [panel 
(b) of Figs.~\ref{fig:ge_e2} and \ref{fig:se_e2}]. The quantity $B(E2; 
0^+_2\rightarrow 2^+_1)$, shown in panel (c) of Figs.~\ref{fig:ge_e2} 
and \ref{fig:se_e2}, can be regarded as a measure of the mixing between 
different intrinsic configurations. The experimental $B(E2;
0^+_2\rightarrow 2^+_1)$ value is very large around $N=38$ or 40 where,  
a pronounced configuration mixing could be expected.  Such a large 
value is not reproduced in our calculations. In this case, the origin 
of the discrepancy between our predictions and the experimental results could be  
associated to a weak mixing between the $2^+_1$ and $0^+_2$ states 
in our model. 
For both the Ge and Se chains, there are some
discrepancies between the predicted and experimental $B(E2;
2^+_2\rightarrow 2^+_1)$ values at the quantitative level [panel (d) of 
Figs.~\ref{fig:ge_e2} and \ref{fig:se_e2}]. 
Nevertheless, the experimental trend, i.e., the $B(E2;
2^+_2\rightarrow 2^+_1)$ transition probability reaches its largest
value at around $N=40$, being almost of the same order of magnitude as
$B(E2; 2^+_1\rightarrow 0^+_1)$, is reproduced rather well by the
present calculation. 
Furthermore, the neutron number $N=40$ is precisely the region where  the 
Gogny-D1M energy surfaces display a pronounced $\gamma$-softness. 

\subsubsection{Spectroscopic quadrupole moments}


\begin{figure}[htb!]
\begin{center}
\includegraphics[width=\columnwidth]{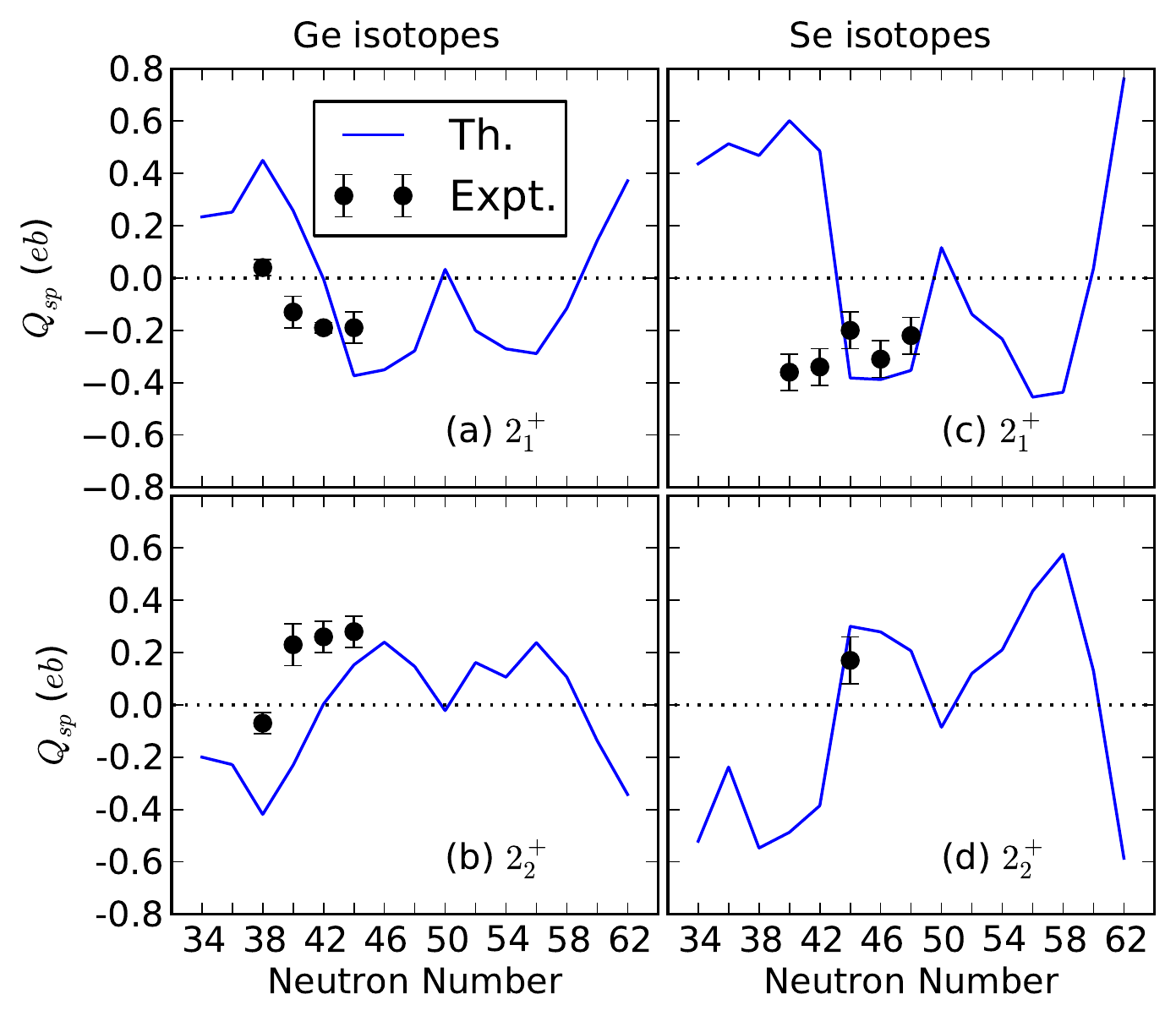}
\caption{(Color online) Spectroscopic quadrupole moments 
$Q_{sp}$ for the $2^+_1$ and $2^+_2$ states in $e$b units. 
The experimental values are taken from Refs.~\cite{data,stone2005}.}
\label{fig:qspec}
\end{center}
\end{figure}

The spectroscopic quadrupole moments $Q_{sp}$ corresponding to the 
$2^+_1$ and $2^+_2$ states in Ge and Se nuclei, are shown in 
Fig.~\ref{fig:qspec} where, they are also compared with the available 
experimental data \cite{data,stone2005}. The predicted postivie spectroscopic 
quadrupole moments $Q_{sp}(2^+_1)$ for $34\le N\le 38$ [panel (a)] 
indicate that the corresponding states are oblate. In our 
calculations, the $2^+_1$ wave functions for  $^{72,74}$Ge are 
dominated by the intruder oblate and triaxial configurations, 
respectively. Consequently, their  $Q_{sp}(2^+_1)$ moments are positive 
and nearly zero, whereas experimentally $Q_{sp}(2^+_1)<0$ at both $N=40$
and 42. 
Similarly, at variance with the data, the predicted $Q_{sp}(2^+_2)$  are negative and
approximately 0 for $^{72}$Ge and $^{74}$Ge, respectively. 
Exception
made of $^{82}$Ge for which $Q_{sp}(2^+_1)\approx 0$, the predicted
spectroscopic quadrupole moments $Q_{sp}(2^+_1)$ are negative for most of the heavier Ge 
isotopes. The sign of the $Q_{sp}(2^+_2)$ values [panel (b)] is the 
opposite to the one of the  $Q_{sp}(2^+_1)$ moments. As can be seen from the figure 
[panels (a) and (b)] our calculations qualitatively follow the experimental trends 
for both $Q_{sp}(2^+_1)$ and $Q_{sp}(2^+_2)$ in Ge isotopes, i.e., 
the decrease (increase) in $Q_{sp}(2^+_1)$ ($Q_{sp}(2^+_2)$) as a
function of $N$ from $N=38$ to 44. 
Similar 
conclusions can be drawn for the predicted $Q_{sp}$ moments in the case 
of Se isotopes [panels (c) and (d)]. In particular, our results for 
$Q_{sp}(2^+_1)$ agree well with the experimental ones at  $N=44$, 46 
and 48 as well as with the only available data on  $Q_{sp}(2^+_2)$  at 
$N=44$. Nevertheless, at variance with the experiment, in our 
calculations $Q_{sp}(2^+_1) > 0$ for $^{74,76}$Se, similarly to their
isotones $^{72,74}$Ge.

\subsubsection{E0 properties}


\begin{figure}[htb!]
\begin{center}
\includegraphics[width=\columnwidth]{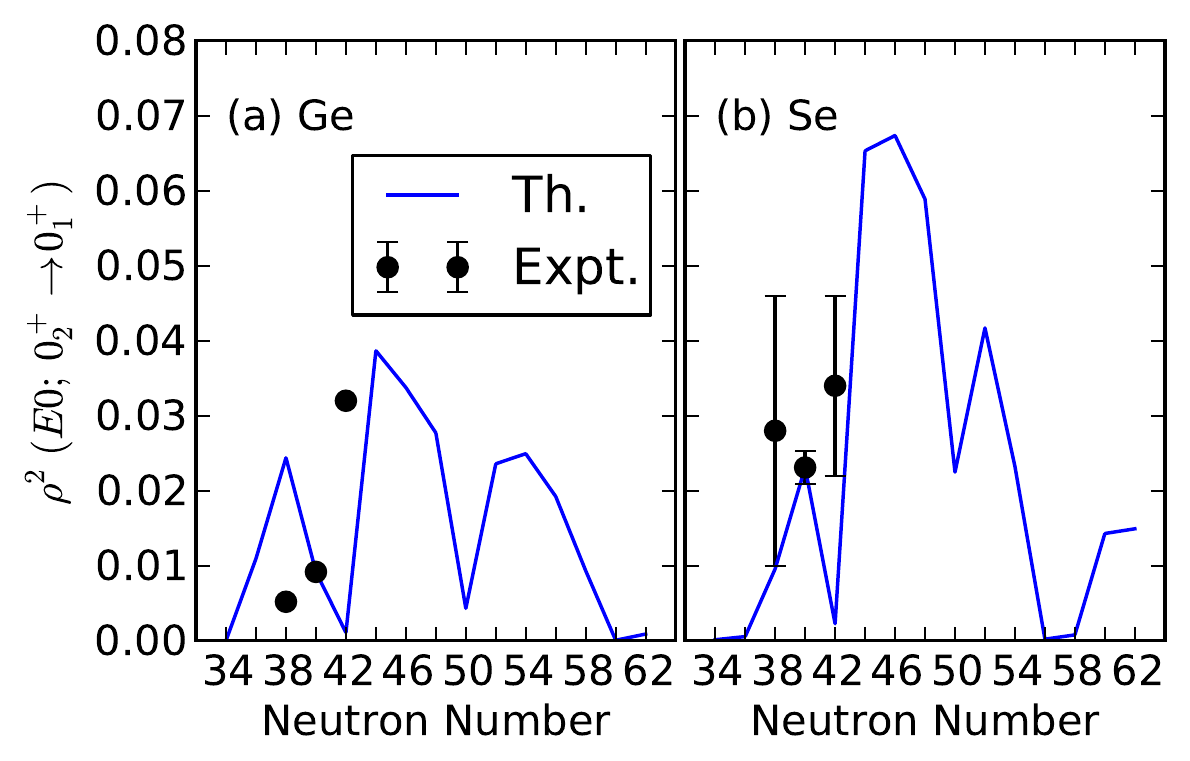}
\caption{(Color online) The $\rho^2(E0; 0^+_2\rightarrow 0^+_1)$ values, 
obtained within the mapped IBM  framework for  Ge and Se nuclei, 
are compared with the available experimental data 
taken from Ref.~\cite{kibedi2005}.}
\label{fig:e0}
\end{center}
\end{figure}

The E0 transition strength values between  $0^+$ states can be regarded as a signature 
of both shape/phase transitions and shape coexistence. The $\rho^2(E0; 
0^+_2\rightarrow 0^+_1)$ values, obtained within the mapped IBM 
framework, for  Ge and Se nuclei are compared with the available 
experimental data  \cite{kibedi2005} in Fig.~\ref{fig:e0}. The peaks 
observed in the predicted $\rho^2(E0; 0^+_2\rightarrow 0^+_1)$ values,
shown in panels (a) and (b) of the figure, characterize the 
structural evolution along both isotopic chains. In the case of the Ge 
isotopes, for example, the peak at $N\approx 38$ can be associated with 
the emergence of shape coexistence while the increase in the 
predicted $\rho^2(E0; 0^+_2\rightarrow 0^+_1)$ values towards $N=44$
suggest the development of quadrupole collectivity in the considered nuclei. For 
both chains $\rho^2(E0; 0^+_2\rightarrow 0^+_1) \approx 0$ at $N=42$, 
considerably underestimating the experimental value. This implies, that 
the mixing between the $0^+_1$ and $0^+_2$ states is too weak in this 
case. As already shown in Fig.~\ref{fig:frac}, for $^{74}$Ge and 
$^{76}$Se, the $0^+_1$ state in the present analysis is made almost
entirely of the intruder ($\gamma$-soft) configuration while the
intruder component is negligible in the  $0^+_2$ state. As can be seen
from panel (b), 
$\rho^2(E0; 0^+_2\rightarrow 0^+_1)$ becomes larger for neutron-rich Se 
isotopes with $N\geq 60$. This indicates the strong mixing between the 
normal and intruder configurations in their $0^+_1$ and $0^+_2$ wave 
functions [see, panel (b) of Fig.~\ref{fig:frac}].

\subsection{Level schemes of selected isotopes\label{sec:detail}}


\begin{figure}
\begin{center}
\includegraphics[width=\columnwidth]{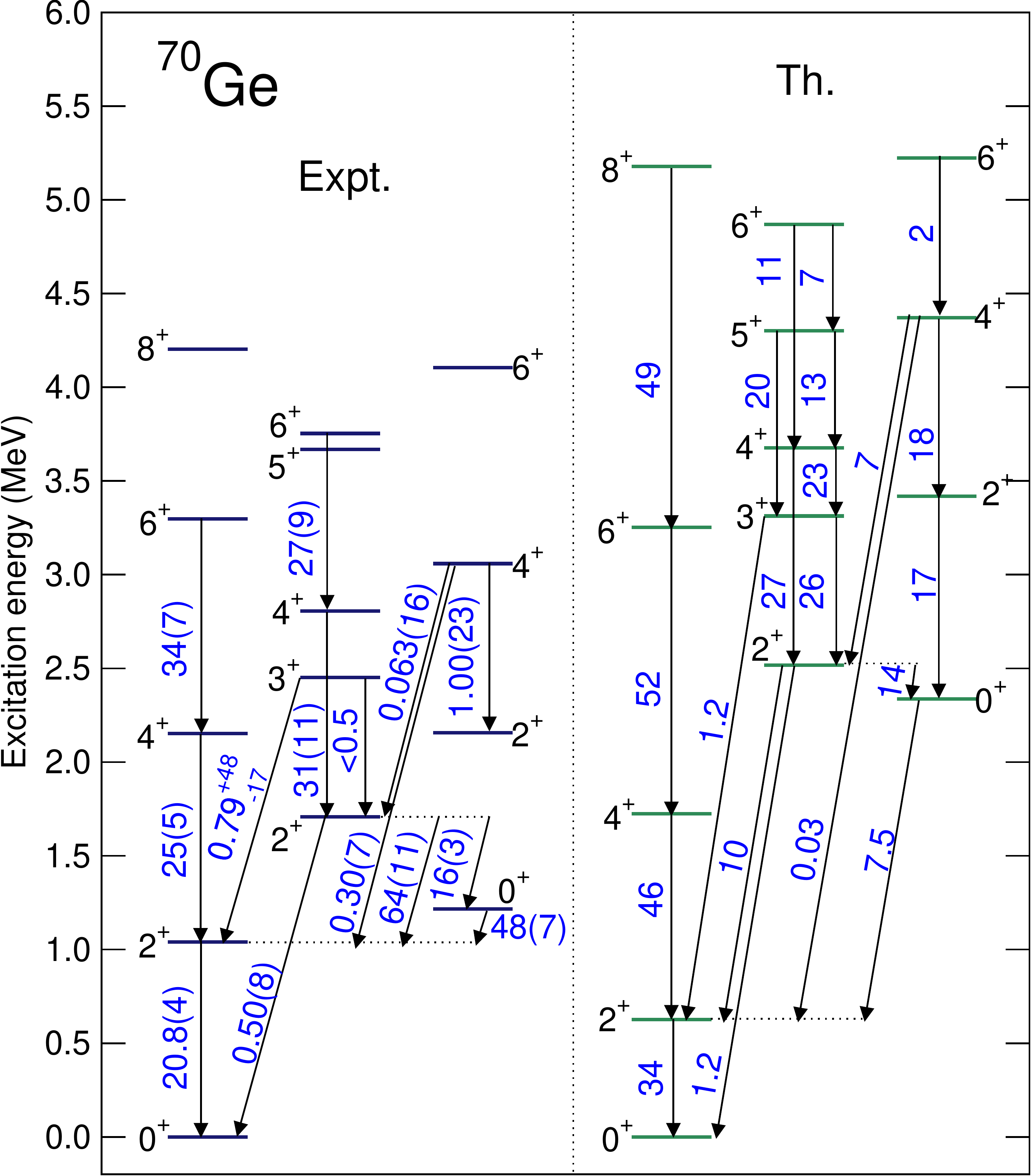}
\caption{(Color online) Low-energy level scheme for $^{70}$Ge. The 
numbers (in blue) near the arrows stand for the $B(E2)$ transition 
strengths in Weisskopf units. Experimental data have been taken from 
Ref.~\cite{data}.}
\label{fig:ge70}
\end{center}
\end{figure}

\begin{figure}
\begin{center}
\includegraphics[width=\columnwidth]{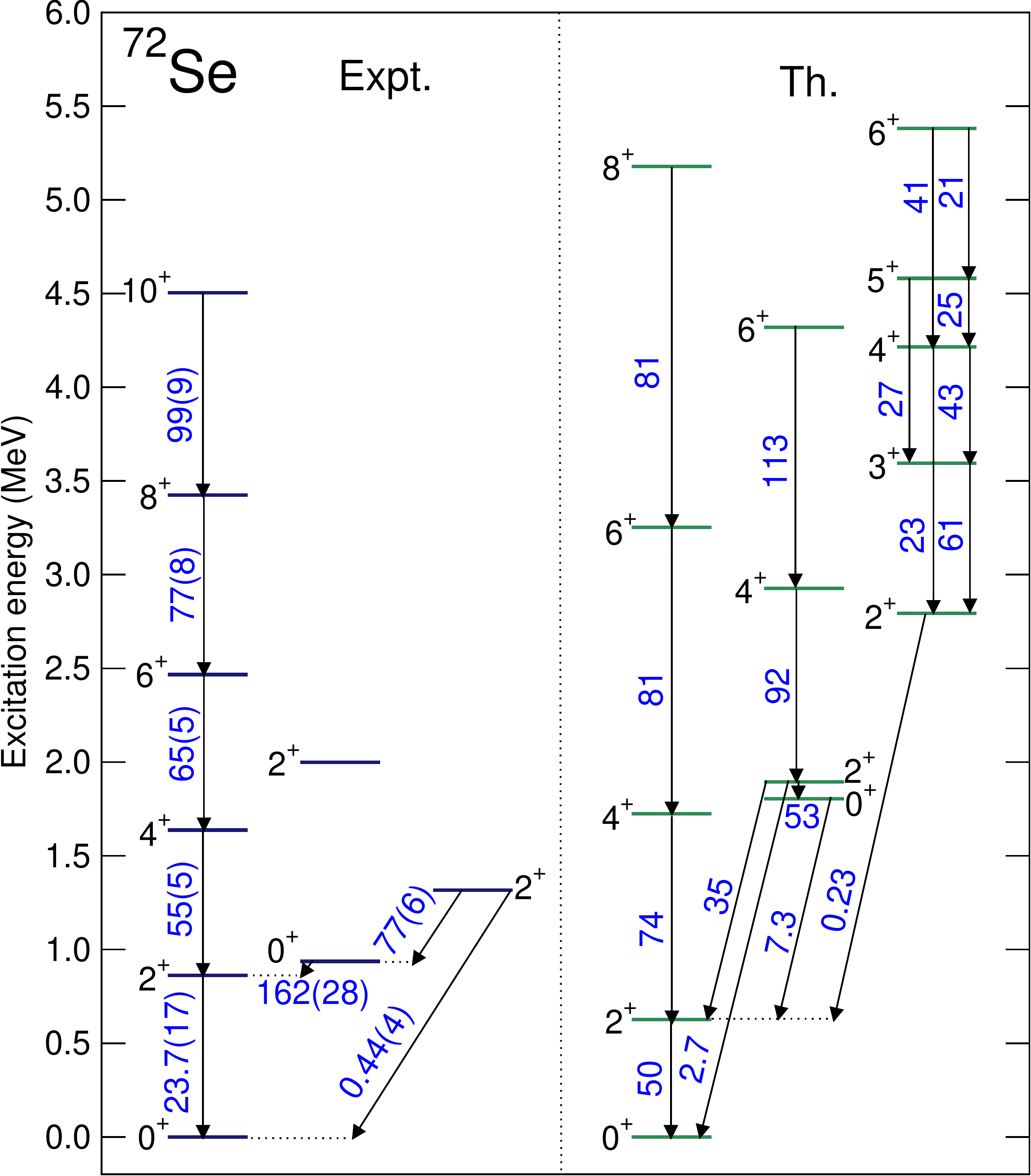}
\caption{(Color online) The same as in Fig.~\ref{fig:ge70}, but for $^{72}$Se.}
\label{fig:se72}
\end{center}
\end{figure}

\begin{figure}
\begin{center}
\includegraphics[width=\columnwidth]{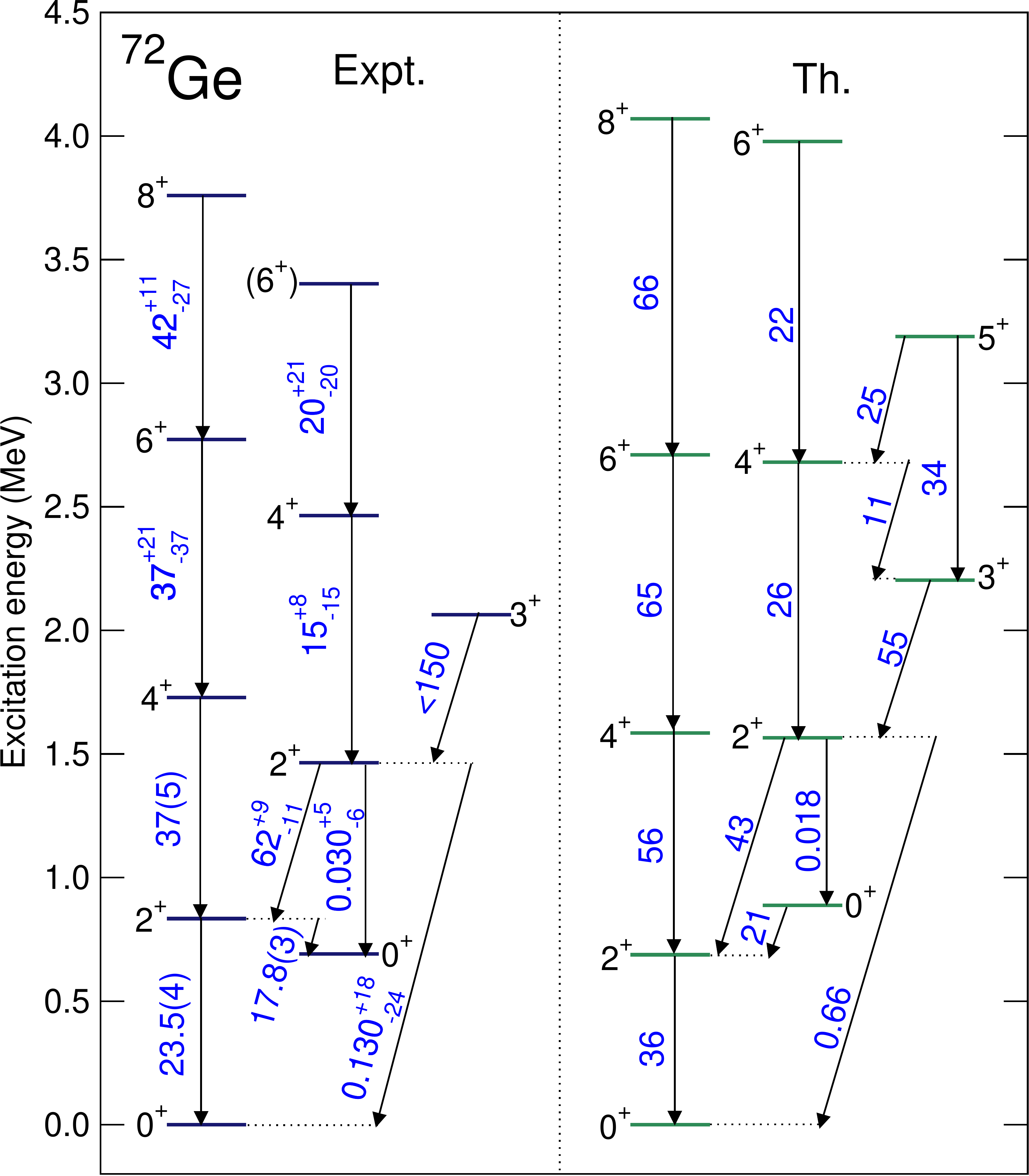}
\caption{(Color online) The same as in Fig.~\ref{fig:ge70}, but for $^{72}$Ge.}
\label{fig:ge72}
\end{center}
\end{figure}

\begin{figure}
\begin{center}
\includegraphics[width=\columnwidth]{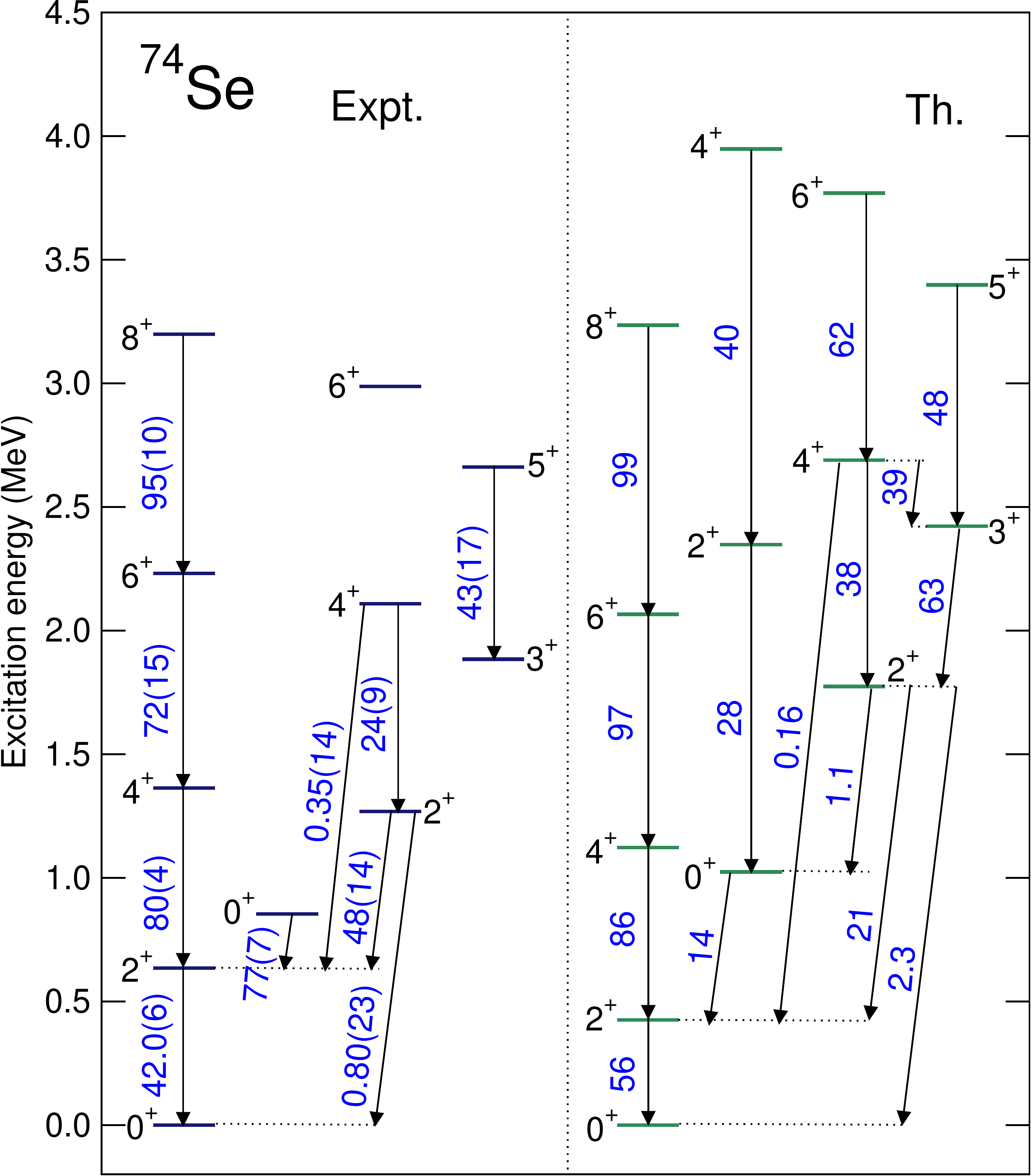}
\caption{(Color online) The same as in Fig.~\ref{fig:ge70}, but for $^{74}$Se.}
\label{fig:se74}
\end{center}
\end{figure}

\begin{figure}
\begin{center}
\includegraphics[width=\columnwidth]{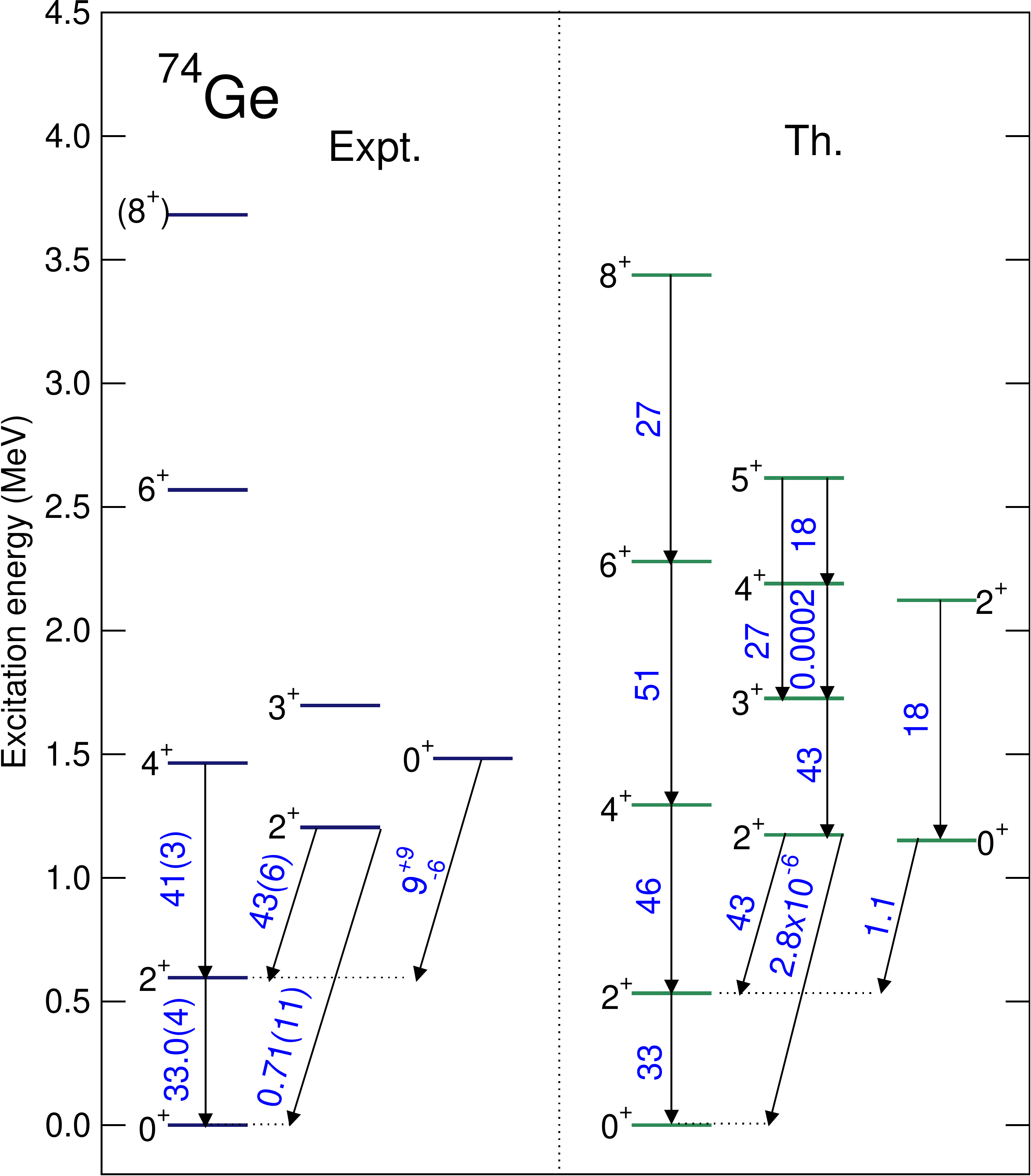}
\caption{(Color online) The same as in Fig.~\ref{fig:ge70}, but for $^{74}$Ge.}
\label{fig:ge74}
\end{center}
\end{figure}

\begin{figure}
\begin{center}
\includegraphics[width=\columnwidth]{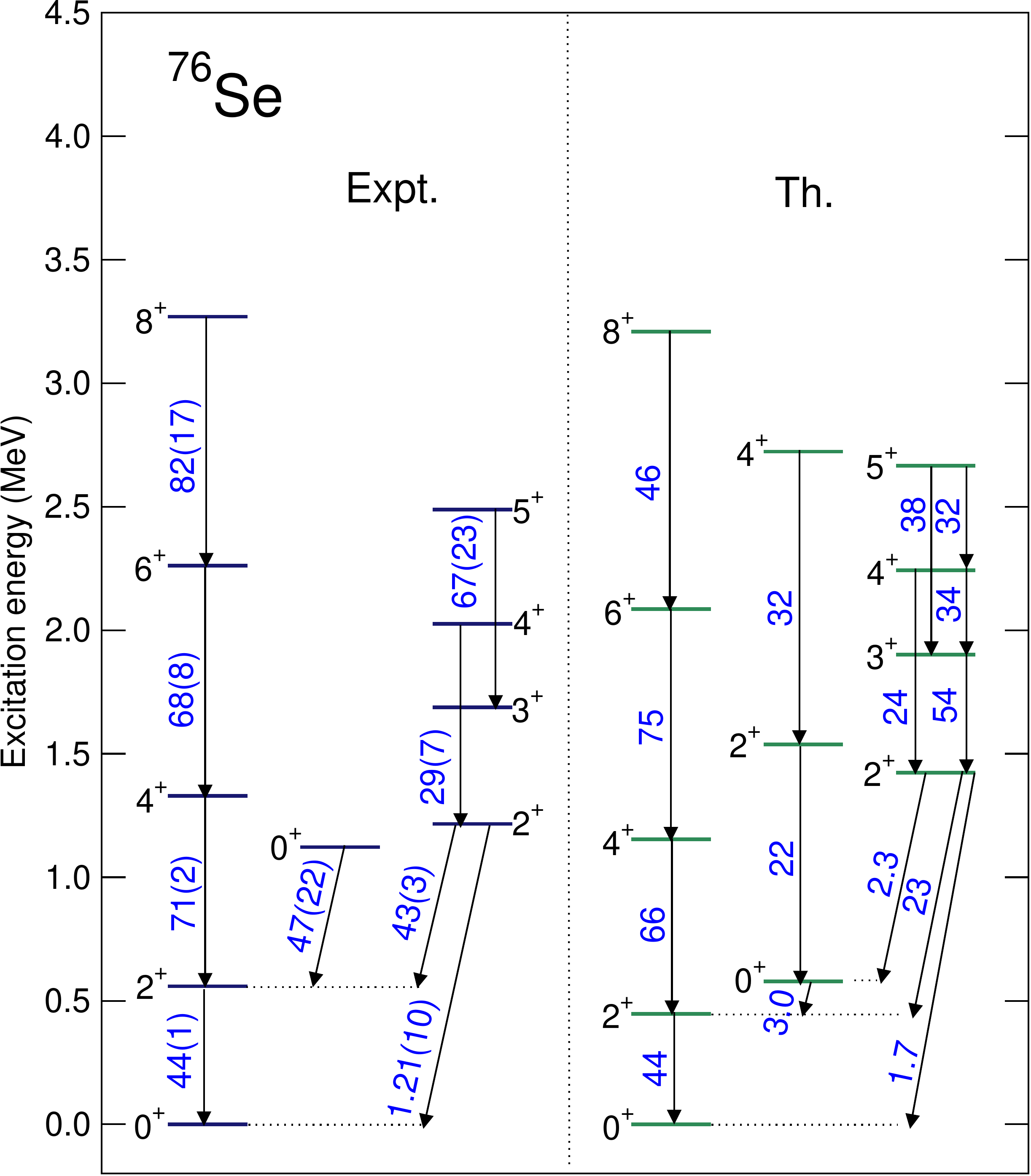}
\caption{(Color online) The same as in Fig.~\ref{fig:ge70}, but for $^{76}$Se.}
\label{fig:se76}
\end{center}
\end{figure}

\begin{figure}
\begin{center}
\includegraphics[width=\columnwidth]{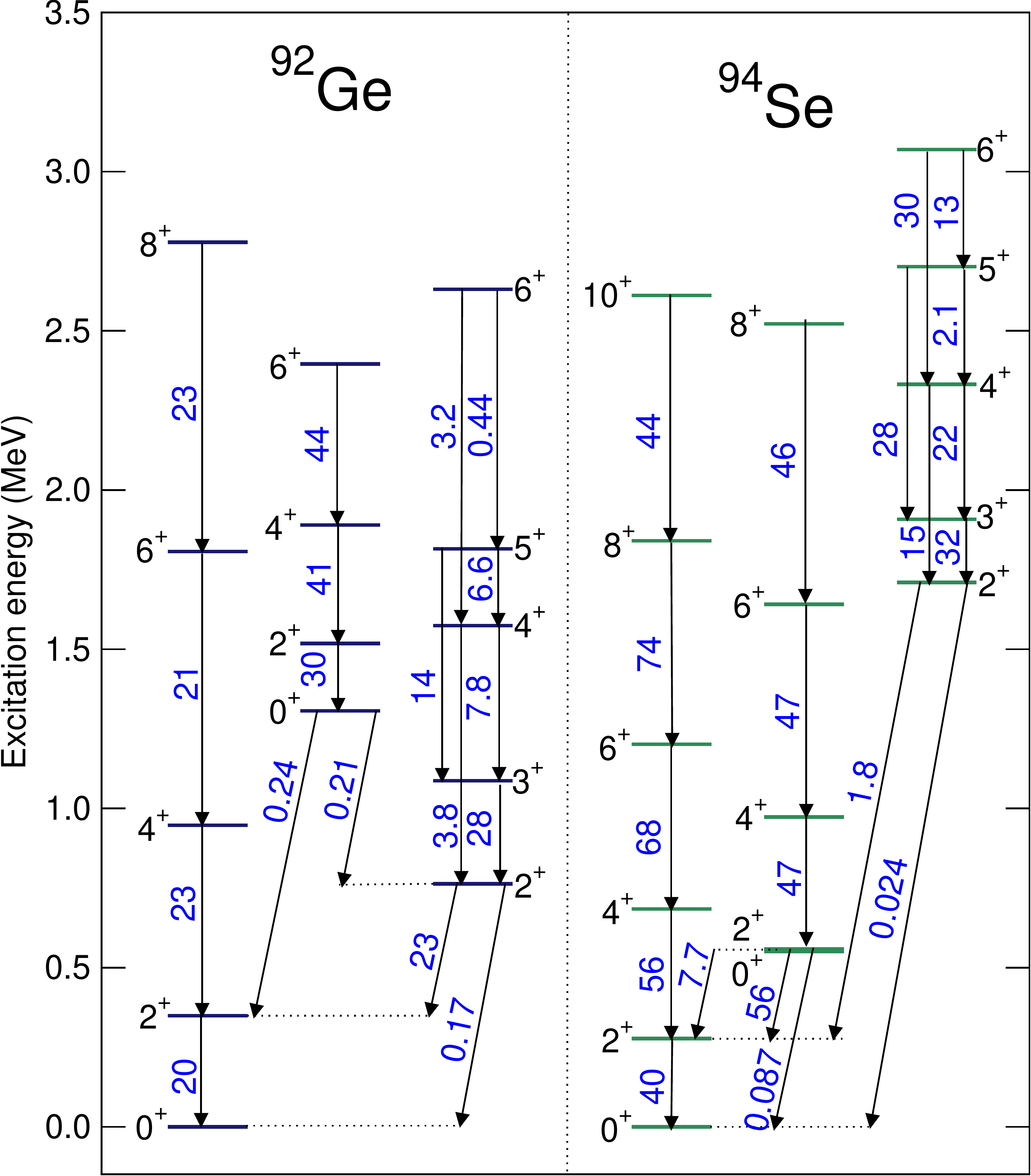}
\caption{(Color online) Low-energy level schemes for the 
$N=60$ isotones $^{92}$Ge and $^{94}$Se. Note that the 
theoretical $0^+_2$ and $2^+_2$ levels of $^{94}$Se are almost degenerated. 
The $B(E2)$ transition strengths relevant to these states are
$B(E2; 0^+_2\rightarrow 2^+_1)=7.7$, 
$B(E2; 2^+_2\rightarrow 2^+_1)=56$, 
$B(E2; 2^+_2\rightarrow 0^+_1)=0.087$, 
$B(E2; 2^+_2\rightarrow 0^+_2)=38$, and 
$B(E2; 4^+_2\rightarrow 2^+_2)=47$ W.u.}
\label{fig:n60}
\end{center}
\end{figure}

In this section, we further demonstrate the ability of our 
fermion-to-boson mapping procedure to describe not only the overall 
systematics of the spectroscopic properties in the studied Ge and Se  
chains but also to account for the detailed band structures and decay 
patterns of individual nuclei in comparison with the experiment. In 
particular, we consider the nuclei $^{70,72,74}$Ge and $^{72,74,76}$Se 
which correspond to an abrupt shape transition and the emergence of 
shape coexistence in their isotopic chains. We will also discuss the 
level schemes obtained for the $N=60$ isotones $^{92}$Ge and $^{94}$Se. 
The level schemes presented in what follows, have been classified into 
bands according to their dominant E2 decays.

\subsubsection{$N=38$ isotones}

The low-energy level schemes obtained for the $N=38$ isotones $^{70}$Ge 
and $^{72}$Se are depicted in Figs.~\ref{fig:ge70} and \ref{fig:se72}. 
The experimental ground-state band exhibits an almost equal spacing 
between its members. On the other hand, the theoretical ground-state 
band, mainly coming from the oblate normal configuration [see, 
Fig.~\ref{fig:frac}], rather looks like a regular collective band 
approximately following the $J(J+1)$ systematics in the rotational limit 
and are more stretched for higher spins. This could be due to the fact
that the Gogny-D1M energy surfaces for these nuclei exhibit a rather
pronounced oblate minimum (see, Figs.~\ref{fig:pes-ge-hfb} and
\ref{fig:pes-se-hfb}), and the resultant mapped Hamiltonian gives more 
collective feature than is suggested experimentally. 
In the case of $^{70}$Ge, our calculations provide the band built on the 
$0^+_2$ state (almost 50\% of the wave function is made of the intruder 
prolate configuration) as well as the quasi-$\gamma$ band with the 
sequence of states ($2^+$, $3^+$, $4^+$, $5^+$, $6^+$, $\ldots$). The 
bandheads of these bands are rather overestimated. Similar excited 
bands are found for $^{72}$Se, but the one built on the $0^+_2$ state 
is lower than the quasi-$\gamma$ band. 

\subsubsection{$N=40$ isotones}

The low-energy level schemes, obtained for the nuclei $^{72}$Ge and 
$^{74}$Se, are compared in Figs.~\ref{fig:ge72} and \ref{fig:se74} with 
the experimental data. For those $N=40$ isotones, the corresponding 
Gogny-D1M energy surfaces exhibit a coexistence between  spherical and 
oblate shapes. From the experimental point of view, the energy of the 
$0^+_2$ state is the lowest precisely at $N=40$. This, together with 
the strong $B(E2; 0^+_2\rightarrow 2^+_1)$ transition and the
$Q_{sp}(2^+)$ values shown in Fig.~\ref{fig:qspec}, suggests a 
pronounced mixing between oblate and prolate configurations in those 
nuclear systems. As can be seen from Figs.~\ref{fig:ge72} and
\ref{fig:se74}, our calculations describe well the experimental spectra
(including the energy of the 
$0^+_2$ states) and B(E2) transition probabilities. However, as already 
discussed in Sec.~\ref{sec:em}, the $B(E2; 0^+_2\rightarrow 2^+_1)$ 
value is underestimated within our model because the mixing between the 
two configurations is not strong enough. Previous calculations, within 
the 5D collective Hamiltonian approach \cite{wang2015}, have also 
provided a reasonable description of the low-energy spectra and decay 
patterns for the same nuclei while overestimating the $0^+_2$ energy  
in $^{72}$Ge.

\subsubsection{$N=42$ isotones}

The spectra obtained for $^{74}$Ge and $^{76}$Se are compared, in 
Figs.\ref{fig:ge74} and \ref{fig:se76}, with the experimental ones. At 
$N=42$, our Gogny-D1M energy surfaces displayed a coexistence between  
spherical and $\gamma$-soft  minima. From the experimental point of 
view, the lower energy of the $2^+_2$ bandhead  of the quasi-$\gamma$ 
band and the strong $B(E2; 2^+_2\rightarrow 2^+_1)$ transition suggest 
that those $N=42$ isotones could be considered as  examples of 
$\gamma$-softness. Our calculations describe reasonably well the 
quasi-$\gamma$ band with the sequence of states ($2^+_{\gamma}, 
3^+_{\gamma}, 4^+_{\gamma}, 5^+_{\gamma}, \ldots$) as well as the 
$B(E2; 2^+_2\rightarrow 2^+_1)$ transition probability which is 
comparatively as large as the $B(E2; 2^+_1\rightarrow 0^+_1)$ value. In 
the case of $^{76}$Se, both the computed and empirical quasi-$\gamma$ 
bands consist of $3^+_{\gamma}$ and $4^+_{\gamma}$ levels close in 
energy following the systematics of the $\gamma$-unstable-rotor model 
of Wilets and Jean \cite{wilets1956}. Furthermore, the ground-state 
bands build on the  $0^+_1$ states obtained for $^{74}$Ge and 
$^{76}$Se, are largely made of the intruder configurations 
corresponding to triaxial ($\gamma\approx 30^{\circ}$) and oblate 
($\gamma=60^{\circ}$) minima in the Gogny-D1M energy surfaces, 
respectively. On the other hand, the $0^+_2$ energy for $^{76}$Se is 
underestimated in the present calculation. 
Another deviation of the predicted level scheme from the experimental one
for $^{76}$Se is the fact that the $B(E2; J\rightarrow J-2)$
($J=2^+,4^+,6^+,8^+$) transition strength in the 
predicted yrast band increases with spin but drops suddenly at $J=8^+$,
while experimentally no such sudden decrease is observed even
considering the experimental uncertainty. 
This reflects a general feature of the IBM \cite{IBM}: due to the finite
boson number the in-band E2 transition strength increases with spin, then
reaches its maximum value at certain spin, and finally decreases. 
For both $^{74}$Ge and $^{76}$Se, the $B(E2; 0^+_2\rightarrow 2^+_1)$
transition probability is largely underestimated in the calculation.
We remind also that the predicted $\rho^2(E0; 0^+_2\rightarrow 0^+_1)$
values for these $N=42$ isotones are too small compared to the
experimental values (see, Fig.~\ref{fig:e0}(a) and (b)). 
Both of these discrepancies have the same origin as the ones in the 
case of the $N=40$ isotones.

\subsubsection{$N=60$ isotones}

The spectra obtained for the neutron-rich nuclei $^{92}$Ge and 
$^{94}$Se are shown in Fig.~\ref{fig:n60}. Within  the Gogny-D1M HFB 
framework, a $\gamma$-soft oblate minimum has been found for the former 
while for the latter our calculations predict prolate-oblate shape 
coexistence. The first excited band for $^{92}$Ge is predicted to be a 
quasi-$\gamma$ band. The large $B(E2; 2^+_2\rightarrow 2^+_1)$ value 
obtained for this nucleus (of the same order of magnitude as the 
in-band $B(E2; 2^+_1\rightarrow 0^+_1)$ strength) indicates a 
pronounced $\gamma$-softness. Furthermore, the $2^+$ and $3^+$ as well 
as the $4^+$ and $5^+$ levels of the quasi-$\gamma$ band are close to 
each other, which is rather consistent with the rigid-triaxial-rotor 
picture of Davydov and Filippov \cite{davydov1958}. In the case of 
$^{94}$Se, the level scheme suggests two coexisting $0^+$ bands (the 
$0^+_1$ ground-state band coming from the oblate normal configuration 
and the $0^+_2$ band coming from the prolate intruder configuration). 
The small $B(E2;0^+_2\rightarrow 2^+_1)$ transition probability 
confirms the weak mixing between oblate and prolate configurations in 
this nucleus. Our calculations also provide a quasi-$\gamma$ band 
that exhibits the rigid-triaxial-rotor feature, being much higher in 
energy than in $^{92}$Ge.


\section{Sensitivity analysis\label{sec:d1md1s}}


\begin{figure}
\begin{center}
\includegraphics[width=\columnwidth]{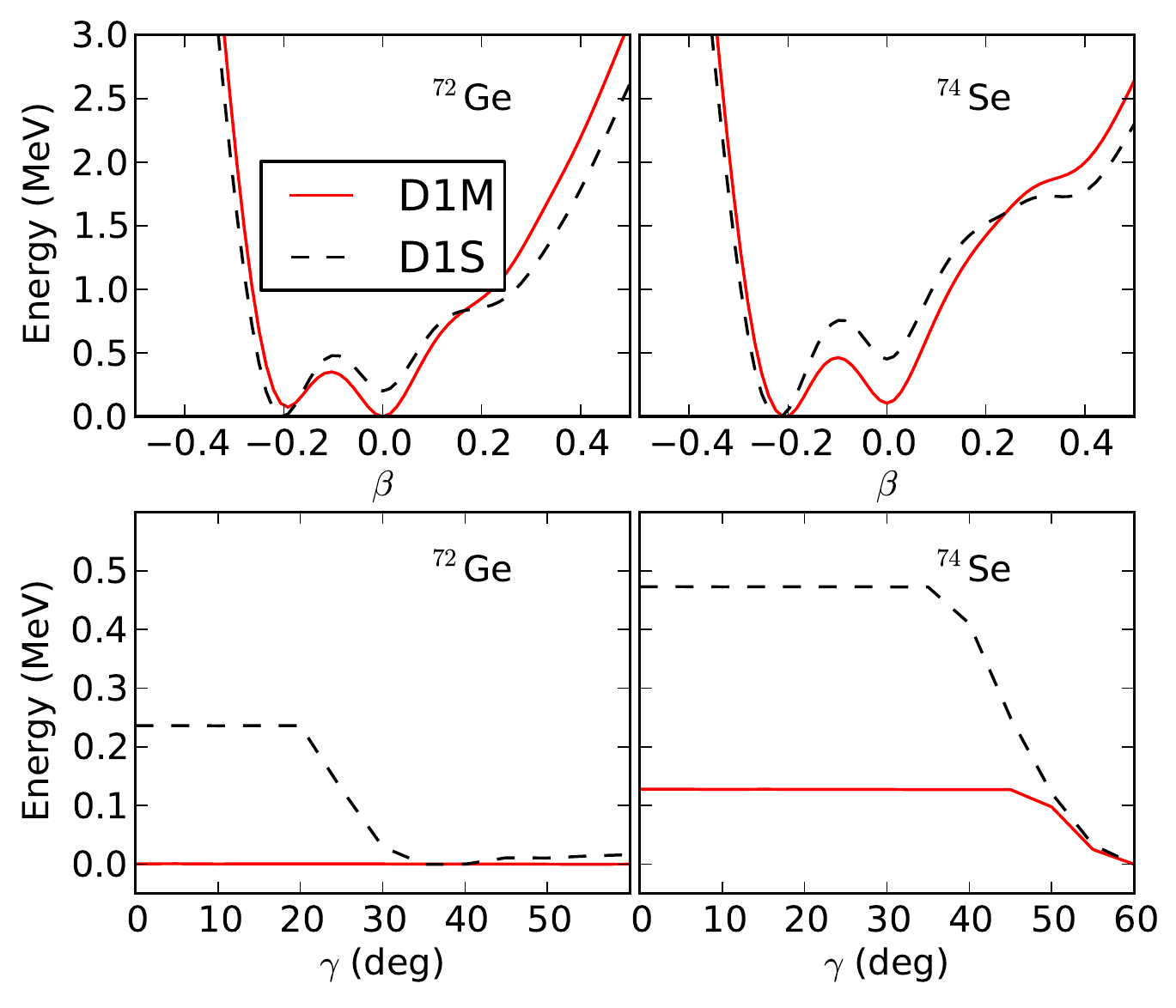}
\caption{(Color online) The Gogny-HFB energy curves for the $N=40$ 
isotones $^{72}$Ge and $^{74}$Se are depicted (upper panels) as functions 
of the axial deformation parameter $\beta$ ($\gamma=0^{\circ}$). In the
lower panels the HFB energies are shown as functions of $\gamma$. For each
$\gamma$ value the parameter $\beta$ is chosen as to minimize the energy. 
Results are shown for both the Gogny-D1S and Gogny-D1M EDFs.}
\label{fig:d1md1s_pes}
\end{center}
\end{figure}

\begin{figure}
\begin{center}
\includegraphics[width=\columnwidth]{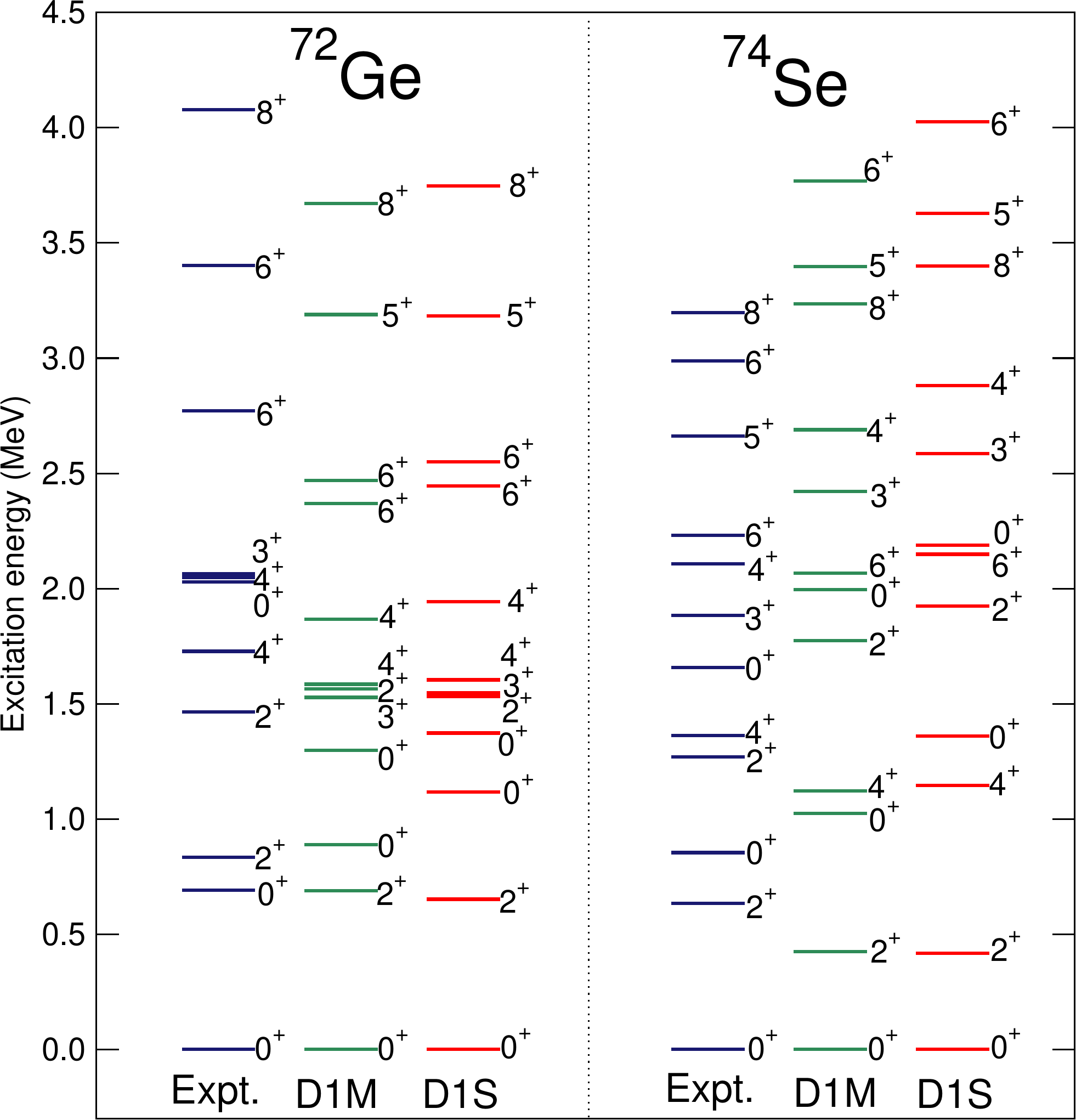}
\caption{(Color online) Low-energy spectra obtained for $^{72}$Ge and $^{74}$Se 
within the fermion-to-boson mapping procedure based on the Gogny-D1S and 
Gogny-D1M EDFs. The experimental spectra are also included to facilitate 
the comparison.}
\label{fig:d1md1s}
\end{center}
\end{figure}

As already pointed out in previous sections, there are several model 
assumptions that could affect our  results for the spectroscopic 
properties of the studied nuclei. In this section, we turn our 
attention to the sensitivity of our results with respect to the 
underlying Gogny-EDF that provides the starting point for our 
fermion-to-boson mapping scheme. To this end, in the upper panels of 
Fig.~\ref{fig:d1md1s_pes} we have plotted the Gogny-HFB energy curves 
for the $N=40$ isotones $^{72}$Ge and $^{74}$Se as functions of the 
axial deformation parameter $\beta$ ($\gamma=0^{\circ}$). In the lower 
panels of the same figure, we have depicted the HFB energies as 
functions of $\gamma$ taking for the parameter $\beta$ the value 
that minimizes the energy for each value of $\gamma$. At the quantitative 
level there are certain differences between the results provided by the 
two functionals. For example, the D1M energy curve for $^{72}$Ge 
exhibits a global spherical minimum while an oblate one is obtained 
with the D1S parametrization. On the other hand, for $^{74}$Se, both 
parameter sets lead to an oblate global minimum but we obtain a softer 
behavior along the $\gamma$ direction with the Gogny-D1M than with the 
Gogny-D1S EDF (see lower panels).

The spectra obtained for $^{72}$Ge and $^{74}$Se with the two Gogny EDFs 
are compared in  Fig.~\ref{fig:d1md1s}. The experimental data are also 
included in the plots to facilitate the comparison. It is satisfying to observe that 
there is no major difference between the spectra provided by both 
parametrizations  of the Gogny-EDF, exception made of the $0^+_2$ 
energy level. Such a difference could be attributed to the different 
topology of the corresponding HFB energy surfaces. We have also  
checked that  the spectroscopic properties obtained for all the 
considered nuclei $^{66-94}$Ge and $^{68-96}$Se with the D1S 
parametrization, are almost identical to the ones obtained with the D1M 
set. This is the reason why we have not discuss them in detail in the 
present paper. 
  

\section{Conclusion\label{sec:summary}}


In this study, we have considered both the shape/phase transitions and 
shape coexistence in the Ge and Se isotopic chains. To this end, 
calculations have been carried out for the nuclei $^{66-94}$Ge and 
$^{68-96}$Se within the Gogny-HFB framework and, subsequently, within 
the mapped IBM approximation. The IBM configuration mixing Hamiltonian, 
with parameters determined through the mapping procedure, has been 
diagonalized and the resulting wave functions have been used to compute 
the spectroscopic properties of the considered nuclei. Though a 
restricted form of the IBM-1 Hamiltonian has been employed, our 
calculations provide a reasonable description of the systematics for 
the low-lying energy spectra and  transition strengths.

The Gogny-D1M energy surfaces predict the coexistence between the 
prolate and oblate shapes in the lightest nuclei in both isotopic 
chains. For shapes around $N=40$ coexistence between spherical and 
$\gamma$-soft shapes is observed. When neutron number increases towards
the $N=50$ shell closure weakly deformed prolate shapes are obtained.
On the other hand, for $52\leq N\leq 62$ a number of nuclei exhibiting 
$\gamma$-soft shapes and coexistence between prolate and oblate shapes are
observed. 
The behaviors of the derived IBM parameters, resulting low-lying energy 
levels, $B(E2)$ transition strengths, spectroscopic quadrupole moments, 
and $\rho^2(E0; 0^+_2\rightarrow 0^+_1)$ values, correlate well with 
the systematic of the Gogny-D1M energy surface. Through the analysis 
of the IBM wave functions, the low-lying $0^+_2$ state around $N=40$ 
has been shown to arise either from the intruder configuration 
associated with the $\gamma$-soft minimum or the normal configuration 
associated with the closely-lying spherical ground state minimum. 
Around this neutron number, our calculation also identifies signatures 
of $\gamma$ softness. On the neutron-rich side with $N\approx 60$, our 
calculation further predicts many examples of the $\gamma$-soft 
spectra and low-lying $0^+_2$ band.

On the other hand, we have also pointed out several discrepancies 
between our results and experimental data. In particular, our 
calculation underestimates the $B(E2;0^+_2\rightarrow 2^+_1)$ 
transition strength for $N\leq 42$, indicating that the mixing between 
the different configurations is too small. This is obviously due to the 
chosen parameters for the IBM Hamiltonian, particularly the too small 
strength parameter of the mixing interaction which could be a consequence of
the topology of the Gogny EDF energy surfaces and/or the assumptions 
made at the IBM level. In this respect, the form of the IBM Hamiltonian 
employed in this study may be too simple, and some additional terms 
could be included in the Hamiltonian. As is well known, the use of the IBM-1 is 
particularly justified for heavy nuclei, where protons and neutrons 
occupy different major shells \cite{OAI}. However, the lightest 
isotopes considered in this work have nearly equal $Z$ and $N$ values and, 
therefore, the presence of proton-neutron pairing effects might not be 
negligible. More realistic calculation should employ versions of the 
IBM that explicitly include isospin degrees of freedom 
\cite{elliot1980,elliot1981}. Nevertheless, these refinements would 
require major extensions of the method and thus present a topic of 
future work.


\acknowledgments
 We would like to thank P. Van Isacker for providing us with the 
computer program IBM-1. K.N. acknowledges support from the Japan 
Society for the Promotion of Science. This work has been supported in 
part by the QuantiXLie Centre of Excellence. The  work of LMR was 
supported by Spanish MINECO grant Nos FPA2015-65929-P and FIS2015-63770-P.

\bibliography{refs}

\begin{thebibliography}{65}%
\makeatletter
\providecommand \@ifxundefined [1]{%
 \@ifx{#1\undefined}
}%
\providecommand \@ifnum [1]{%
 \ifnum #1\expandafter \@firstoftwo
 \else \expandafter \@secondoftwo
 \fi
}%
\providecommand \@ifx [1]{%
 \ifx #1\expandafter \@firstoftwo
 \else \expandafter \@secondoftwo
 \fi
}%
\providecommand \natexlab [1]{#1}%
\providecommand \enquote  [1]{``#1''}%
\providecommand \bibnamefont  [1]{#1}%
\providecommand \bibfnamefont [1]{#1}%
\providecommand \citenamefont [1]{#1}%
\providecommand \href@noop [0]{\@secondoftwo}%
\providecommand \href [0]{\begingroup \@sanitize@url \@href}%
\providecommand \@href[1]{\@@startlink{#1}\@@href}%
\providecommand \@@href[1]{\endgroup#1\@@endlink}%
\providecommand \@sanitize@url [0]{\catcode `\\12\catcode `\$12\catcode
  `\&12\catcode `\#12\catcode `\^12\catcode `\_12\catcode `\%12\relax}%
\providecommand \@@startlink[1]{}%
\providecommand \@@endlink[0]{}%
\providecommand \url  [0]{\begingroup\@sanitize@url \@url }%
\providecommand \@url [1]{\endgroup\@href {#1}{\urlprefix }}%
\providecommand \urlprefix  [0]{URL }%
\providecommand \Eprint [0]{\href }%
\providecommand \doibase [0]{http://dx.doi.org/}%
\providecommand \selectlanguage [0]{\@gobble}%
\providecommand \bibinfo  [0]{\@secondoftwo}%
\providecommand \bibfield  [0]{\@secondoftwo}%
\providecommand \translation [1]{[#1]}%
\providecommand \BibitemOpen [0]{}%
\providecommand \bibitemStop [0]{}%
\providecommand \bibitemNoStop [0]{.\EOS\space}%
\providecommand \EOS [0]{\spacefactor3000\relax}%
\providecommand \BibitemShut  [1]{\csname bibitem#1\endcsname}%
\let\auto@bib@innerbib\@empty
\bibitem [{\citenamefont {Bohr}\ and\ \citenamefont {Mottelsson}(1975)}]{BM}%
  \BibitemOpen
  \bibfield  {author} {\bibinfo {author} {\bibfnamefont {A.}~\bibnamefont
  {Bohr}}\ and\ \bibinfo {author} {\bibfnamefont {B.~M.}\ \bibnamefont
  {Mottelsson}},\ }\href@noop {} {\emph {\bibinfo {title} {Nuclear
  Structure}}},\ Vol.~\bibinfo {volume} {2}\ (\bibinfo  {publisher} {Benjamin,
  New York, USA},\ \bibinfo {year} {1975})\ p.~\bibinfo {pages}
  {45}\BibitemShut {NoStop}%
\bibitem [{\citenamefont {Casten}(2005)}]{CasBook}%
  \BibitemOpen
  \bibfield  {author} {\bibinfo {author} {\bibfnamefont {R.~F.}\ \bibnamefont
  {Casten}},\ }\href@noop {} {\emph {\bibinfo {title} {Nuclear Structure from a
  Simple Perspective}}}\ (\bibinfo  {publisher} {Oxford University Press,
  Oxford, England},\ \bibinfo {year} {2005})\BibitemShut {NoStop}%
\bibitem [{\citenamefont {Cejnar}\ \emph {et~al.}(2010)\citenamefont {Cejnar},
  \citenamefont {Jolie},\ and\ \citenamefont {Casten}}]{cejnar2010}%
  \BibitemOpen
  \bibfield  {author} {\bibinfo {author} {\bibfnamefont {P.}~\bibnamefont
  {Cejnar}}, \bibinfo {author} {\bibfnamefont {J.}~\bibnamefont {Jolie}}, \
  and\ \bibinfo {author} {\bibfnamefont {R.~F.}\ \bibnamefont {Casten}},\
  }\href {\doibase 10.1103/RevModPhys.82.2155} {\bibfield  {journal} {\bibinfo
  {journal} {Rev. Mod. Phys.}\ }\textbf {\bibinfo {volume} {82}},\ \bibinfo
  {pages} {2155} (\bibinfo {year} {2010})}\BibitemShut {NoStop}%
\bibitem [{\citenamefont {G\"urdal}\ \emph {et~al.}(2013)\citenamefont
  {G\"urdal}, \citenamefont {Stefanova}, \citenamefont {Boutachkov},
  \citenamefont {Torres}, \citenamefont {Kumbartzki}, \citenamefont
  {Benczer-Koller}, \citenamefont {Sharon}, \citenamefont {Zamick},
  \citenamefont {Robinson}, \citenamefont {Ahn}, \citenamefont {Anagnostatou},
  \citenamefont {Bernards}, \citenamefont {Elvers}, \citenamefont {Heinz},
  \citenamefont {Ilie}, \citenamefont {Radeck}, \citenamefont {Savran},
  \citenamefont {Werner},\ and\ \citenamefont {Williams}}]{gurdal2013}%
  \BibitemOpen
  \bibfield  {author} {\bibinfo {author} {\bibfnamefont {G.}~\bibnamefont
  {G\"urdal}}, \bibinfo {author} {\bibfnamefont {E.~A.}\ \bibnamefont
  {Stefanova}}, \bibinfo {author} {\bibfnamefont {P.}~\bibnamefont
  {Boutachkov}}, \bibinfo {author} {\bibfnamefont {D.~A.}\ \bibnamefont
  {Torres}}, \bibinfo {author} {\bibfnamefont {G.~J.}\ \bibnamefont
  {Kumbartzki}}, \bibinfo {author} {\bibfnamefont {N.}~\bibnamefont
  {Benczer-Koller}}, \bibinfo {author} {\bibfnamefont {Y.~Y.}\ \bibnamefont
  {Sharon}}, \bibinfo {author} {\bibfnamefont {L.}~\bibnamefont {Zamick}},
  \bibinfo {author} {\bibfnamefont {S.~J.~Q.}\ \bibnamefont {Robinson}},
  \bibinfo {author} {\bibfnamefont {T.}~\bibnamefont {Ahn}}, \bibinfo {author}
  {\bibfnamefont {V.}~\bibnamefont {Anagnostatou}}, \bibinfo {author}
  {\bibfnamefont {C.}~\bibnamefont {Bernards}}, \bibinfo {author}
  {\bibfnamefont {M.}~\bibnamefont {Elvers}}, \bibinfo {author} {\bibfnamefont
  {A.}~\bibnamefont {Heinz}}, \bibinfo {author} {\bibfnamefont
  {G.}~\bibnamefont {Ilie}}, \bibinfo {author} {\bibfnamefont {D.}~\bibnamefont
  {Radeck}}, \bibinfo {author} {\bibfnamefont {D.}~\bibnamefont {Savran}},
  \bibinfo {author} {\bibfnamefont {V.}~\bibnamefont {Werner}}, \ and\ \bibinfo
  {author} {\bibfnamefont {E.}~\bibnamefont {Williams}},\ }\href {\doibase
  10.1103/PhysRevC.88.014301} {\bibfield  {journal} {\bibinfo  {journal} {Phys.
  Rev. C}\ }\textbf {\bibinfo {volume} {88}},\ \bibinfo {pages} {014301}
  (\bibinfo {year} {2013})}\BibitemShut {NoStop}%
\bibitem [{\citenamefont {Corsi}\ \emph {et~al.}(2013)\citenamefont {Corsi},
  \citenamefont {Delaroche}, \citenamefont {Obertelli}, \citenamefont
  {Baugher}, \citenamefont {Bazin}, \citenamefont {Boissinot}, \citenamefont
  {Flavigny}, \citenamefont {Gade}, \citenamefont {Girod}, \citenamefont
  {Glasmacher}, \citenamefont {Grinyer}, \citenamefont {Korten}, \citenamefont
  {Libert}, \citenamefont {Ljungvall}, \citenamefont {McDaniel}, \citenamefont
  {Ratkiewicz}, \citenamefont {Signoracci}, \citenamefont {Stroberg},
  \citenamefont {Sulignano},\ and\ \citenamefont {Weisshaar}}]{corsi2013}%
  \BibitemOpen
  \bibfield  {author} {\bibinfo {author} {\bibfnamefont {A.}~\bibnamefont
  {Corsi}}, \bibinfo {author} {\bibfnamefont {J.-P.}\ \bibnamefont
  {Delaroche}}, \bibinfo {author} {\bibfnamefont {A.}~\bibnamefont
  {Obertelli}}, \bibinfo {author} {\bibfnamefont {T.}~\bibnamefont {Baugher}},
  \bibinfo {author} {\bibfnamefont {D.}~\bibnamefont {Bazin}}, \bibinfo
  {author} {\bibfnamefont {S.}~\bibnamefont {Boissinot}}, \bibinfo {author}
  {\bibfnamefont {F.}~\bibnamefont {Flavigny}}, \bibinfo {author}
  {\bibfnamefont {A.}~\bibnamefont {Gade}}, \bibinfo {author} {\bibfnamefont
  {M.}~\bibnamefont {Girod}}, \bibinfo {author} {\bibfnamefont
  {T.}~\bibnamefont {Glasmacher}}, \bibinfo {author} {\bibfnamefont {G.~F.}\
  \bibnamefont {Grinyer}}, \bibinfo {author} {\bibfnamefont {W.}~\bibnamefont
  {Korten}}, \bibinfo {author} {\bibfnamefont {J.}~\bibnamefont {Libert}},
  \bibinfo {author} {\bibfnamefont {J.}~\bibnamefont {Ljungvall}}, \bibinfo
  {author} {\bibfnamefont {S.}~\bibnamefont {McDaniel}}, \bibinfo {author}
  {\bibfnamefont {A.}~\bibnamefont {Ratkiewicz}}, \bibinfo {author}
  {\bibfnamefont {A.}~\bibnamefont {Signoracci}}, \bibinfo {author}
  {\bibfnamefont {R.}~\bibnamefont {Stroberg}}, \bibinfo {author}
  {\bibfnamefont {B.}~\bibnamefont {Sulignano}}, \ and\ \bibinfo {author}
  {\bibfnamefont {D.}~\bibnamefont {Weisshaar}},\ }\href {\doibase
  10.1103/PhysRevC.88.044311} {\bibfield  {journal} {\bibinfo  {journal} {Phys.
  Rev. C}\ }\textbf {\bibinfo {volume} {88}},\ \bibinfo {pages} {044311}
  (\bibinfo {year} {2013})}\BibitemShut {NoStop}%
\bibitem [{\citenamefont {Toh}\ \emph {et~al.}(2013)\citenamefont {Toh},
  \citenamefont {Chiara}, \citenamefont {McCutchan}, \citenamefont {Walters},
  \citenamefont {Janssens}, \citenamefont {Carpenter}, \citenamefont {Zhu},
  \citenamefont {Broda}, \citenamefont {Fornal}, \citenamefont {Kay},
  \citenamefont {Kondev}, \citenamefont {Kr\'olas}, \citenamefont {Lauritsen},
  \citenamefont {Lister}, \citenamefont {Paw\l{}at}, \citenamefont
  {Seweryniak}, \citenamefont {Stefanescu}, \citenamefont {Stone},
  \citenamefont {Wrzesi\ifmmode~\acute{n}\else \'{n}\fi{}ski}, \citenamefont
  {Higashiyama},\ and\ \citenamefont {Yoshinaga}}]{toh2013}%
  \BibitemOpen
  \bibfield  {author} {\bibinfo {author} {\bibfnamefont {Y.}~\bibnamefont
  {Toh}}, \bibinfo {author} {\bibfnamefont {C.~J.}\ \bibnamefont {Chiara}},
  \bibinfo {author} {\bibfnamefont {E.~A.}\ \bibnamefont {McCutchan}}, \bibinfo
  {author} {\bibfnamefont {W.~B.}\ \bibnamefont {Walters}}, \bibinfo {author}
  {\bibfnamefont {R.~V.~F.}\ \bibnamefont {Janssens}}, \bibinfo {author}
  {\bibfnamefont {M.~P.}\ \bibnamefont {Carpenter}}, \bibinfo {author}
  {\bibfnamefont {S.}~\bibnamefont {Zhu}}, \bibinfo {author} {\bibfnamefont
  {R.}~\bibnamefont {Broda}}, \bibinfo {author} {\bibfnamefont
  {B.}~\bibnamefont {Fornal}}, \bibinfo {author} {\bibfnamefont {B.~P.}\
  \bibnamefont {Kay}}, \bibinfo {author} {\bibfnamefont {F.~G.}\ \bibnamefont
  {Kondev}}, \bibinfo {author} {\bibfnamefont {W.}~\bibnamefont {Kr\'olas}},
  \bibinfo {author} {\bibfnamefont {T.}~\bibnamefont {Lauritsen}}, \bibinfo
  {author} {\bibfnamefont {C.~J.}\ \bibnamefont {Lister}}, \bibinfo {author}
  {\bibfnamefont {T.}~\bibnamefont {Paw\l{}at}}, \bibinfo {author}
  {\bibfnamefont {D.}~\bibnamefont {Seweryniak}}, \bibinfo {author}
  {\bibfnamefont {I.}~\bibnamefont {Stefanescu}}, \bibinfo {author}
  {\bibfnamefont {N.~J.}\ \bibnamefont {Stone}}, \bibinfo {author}
  {\bibfnamefont {J.}~\bibnamefont {Wrzesi\ifmmode~\acute{n}\else
  \'{n}\fi{}ski}}, \bibinfo {author} {\bibfnamefont {K.}~\bibnamefont
  {Higashiyama}}, \ and\ \bibinfo {author} {\bibfnamefont {N.}~\bibnamefont
  {Yoshinaga}},\ }\href {\doibase 10.1103/PhysRevC.87.041304} {\bibfield
  {journal} {\bibinfo  {journal} {Phys. Rev. C}\ }\textbf {\bibinfo {volume}
  {87}},\ \bibinfo {pages} {041304} (\bibinfo {year} {2013})}\BibitemShut
  {NoStop}%
\bibitem [{\citenamefont {Sun}\ \emph {et~al.}(2014)\citenamefont {Sun},
  \citenamefont {Shi}, \citenamefont {Li}, \citenamefont {Hua}, \citenamefont
  {Xu}, \citenamefont {Chen}, \citenamefont {Zhang}, \citenamefont {Song},
  \citenamefont {Meng}, \citenamefont {Wu}, \citenamefont {Hu}, \citenamefont
  {Zhang}, \citenamefont {Liang}, \citenamefont {Xu}, \citenamefont {Li},
  \citenamefont {Li}, \citenamefont {He}, \citenamefont {Zheng}, \citenamefont
  {Ye}, \citenamefont {Jiang}, \citenamefont {Cheng}, \citenamefont {He},
  \citenamefont {Han}, \citenamefont {Li}, \citenamefont {Li}, \citenamefont
  {Li}, \citenamefont {Wang}, \citenamefont {Liu}, \citenamefont {Wu},
  \citenamefont {Luo}, \citenamefont {Yao}, \citenamefont {Yu}, \citenamefont
  {Cao},\ and\ \citenamefont {Sun}}]{sun2014}%
  \BibitemOpen
  \bibfield  {author} {\bibinfo {author} {\bibfnamefont {J.}~\bibnamefont
  {Sun}}, \bibinfo {author} {\bibfnamefont {Z.}~\bibnamefont {Shi}}, \bibinfo
  {author} {\bibfnamefont {X.}~\bibnamefont {Li}}, \bibinfo {author}
  {\bibfnamefont {H.}~\bibnamefont {Hua}}, \bibinfo {author} {\bibfnamefont
  {C.}~\bibnamefont {Xu}}, \bibinfo {author} {\bibfnamefont {Q.}~\bibnamefont
  {Chen}}, \bibinfo {author} {\bibfnamefont {S.}~\bibnamefont {Zhang}},
  \bibinfo {author} {\bibfnamefont {C.}~\bibnamefont {Song}}, \bibinfo {author}
  {\bibfnamefont {J.}~\bibnamefont {Meng}}, \bibinfo {author} {\bibfnamefont
  {X.}~\bibnamefont {Wu}}, \bibinfo {author} {\bibfnamefont {S.}~\bibnamefont
  {Hu}}, \bibinfo {author} {\bibfnamefont {H.}~\bibnamefont {Zhang}}, \bibinfo
  {author} {\bibfnamefont {W.}~\bibnamefont {Liang}}, \bibinfo {author}
  {\bibfnamefont {F.}~\bibnamefont {Xu}}, \bibinfo {author} {\bibfnamefont
  {Z.}~\bibnamefont {Li}}, \bibinfo {author} {\bibfnamefont {G.}~\bibnamefont
  {Li}}, \bibinfo {author} {\bibfnamefont {C.}~\bibnamefont {He}}, \bibinfo
  {author} {\bibfnamefont {Y.}~\bibnamefont {Zheng}}, \bibinfo {author}
  {\bibfnamefont {Y.}~\bibnamefont {Ye}}, \bibinfo {author} {\bibfnamefont
  {D.}~\bibnamefont {Jiang}}, \bibinfo {author} {\bibfnamefont
  {Y.}~\bibnamefont {Cheng}}, \bibinfo {author} {\bibfnamefont
  {C.}~\bibnamefont {He}}, \bibinfo {author} {\bibfnamefont {R.}~\bibnamefont
  {Han}}, \bibinfo {author} {\bibfnamefont {Z.}~\bibnamefont {Li}}, \bibinfo
  {author} {\bibfnamefont {C.}~\bibnamefont {Li}}, \bibinfo {author}
  {\bibfnamefont {H.}~\bibnamefont {Li}}, \bibinfo {author} {\bibfnamefont
  {J.}~\bibnamefont {Wang}}, \bibinfo {author} {\bibfnamefont {J.}~\bibnamefont
  {Liu}}, \bibinfo {author} {\bibfnamefont {Y.}~\bibnamefont {Wu}}, \bibinfo
  {author} {\bibfnamefont {P.}~\bibnamefont {Luo}}, \bibinfo {author}
  {\bibfnamefont {S.}~\bibnamefont {Yao}}, \bibinfo {author} {\bibfnamefont
  {B.}~\bibnamefont {Yu}}, \bibinfo {author} {\bibfnamefont {X.}~\bibnamefont
  {Cao}}, \ and\ \bibinfo {author} {\bibfnamefont {H.}~\bibnamefont {Sun}},\
  }\href {\doibase http://dx.doi.org/10.1016/j.physletb.2014.05.069} {\bibfield
   {journal} {\bibinfo  {journal} {Physics Letters B}\ }\textbf {\bibinfo
  {volume} {734}},\ \bibinfo {pages} {308 } (\bibinfo {year}
  {2014})}\BibitemShut {NoStop}%
\bibitem [{\citenamefont {Yoshinaga}\ \emph {et~al.}(2008)\citenamefont
  {Yoshinaga}, \citenamefont {Higashiyama},\ and\ \citenamefont
  {Regan}}]{yoshinaga2008}%
  \BibitemOpen
  \bibfield  {author} {\bibinfo {author} {\bibfnamefont {N.}~\bibnamefont
  {Yoshinaga}}, \bibinfo {author} {\bibfnamefont {K.}~\bibnamefont
  {Higashiyama}}, \ and\ \bibinfo {author} {\bibfnamefont {P.~H.}\ \bibnamefont
  {Regan}},\ }\href {\doibase 10.1103/PhysRevC.78.044320} {\bibfield  {journal}
  {\bibinfo  {journal} {Phys. Rev. C}\ }\textbf {\bibinfo {volume} {78}},\
  \bibinfo {pages} {044320} (\bibinfo {year} {2008})}\BibitemShut {NoStop}%
\bibitem [{\citenamefont {Honma}\ \emph {et~al.}(2009)\citenamefont {Honma},
  \citenamefont {Otsuka}, \citenamefont {Mizusaki},\ and\ \citenamefont
  {Hjorth-Jensen}}]{honma2009}%
  \BibitemOpen
  \bibfield  {author} {\bibinfo {author} {\bibfnamefont {M.}~\bibnamefont
  {Honma}}, \bibinfo {author} {\bibfnamefont {T.}~\bibnamefont {Otsuka}},
  \bibinfo {author} {\bibfnamefont {T.}~\bibnamefont {Mizusaki}}, \ and\
  \bibinfo {author} {\bibfnamefont {M.}~\bibnamefont {Hjorth-Jensen}},\ }\href
  {\doibase 10.1103/PhysRevC.80.064323} {\bibfield  {journal} {\bibinfo
  {journal} {Phys. Rev. C}\ }\textbf {\bibinfo {volume} {80}},\ \bibinfo
  {pages} {064323} (\bibinfo {year} {2009})}\BibitemShut {NoStop}%
\bibitem [{\citenamefont {Kaneko}\ \emph {et~al.}(2015)\citenamefont {Kaneko},
  \citenamefont {Mizusaki}, \citenamefont {Sun},\ and\ \citenamefont
  {Tazaki}}]{kaneko2015}%
  \BibitemOpen
  \bibfield  {author} {\bibinfo {author} {\bibfnamefont {K.}~\bibnamefont
  {Kaneko}}, \bibinfo {author} {\bibfnamefont {T.}~\bibnamefont {Mizusaki}},
  \bibinfo {author} {\bibfnamefont {Y.}~\bibnamefont {Sun}}, \ and\ \bibinfo
  {author} {\bibfnamefont {S.}~\bibnamefont {Tazaki}},\ }\href {\doibase
  10.1103/PhysRevC.92.044331} {\bibfield  {journal} {\bibinfo  {journal} {Phys.
  Rev. C}\ }\textbf {\bibinfo {volume} {92}},\ \bibinfo {pages} {044331}
  (\bibinfo {year} {2015})}\BibitemShut {NoStop}%
\bibitem [{\citenamefont {Gaudefroy}\ \emph {et~al.}(2009)\citenamefont
  {Gaudefroy}, \citenamefont {Obertelli}, \citenamefont {P\'eru}, \citenamefont
  {Pillet}, \citenamefont {Hilaire}, \citenamefont {Delaroche}, \citenamefont
  {Girod},\ and\ \citenamefont {Libert}}]{gaudefroy2009}%
  \BibitemOpen
  \bibfield  {author} {\bibinfo {author} {\bibfnamefont {L.}~\bibnamefont
  {Gaudefroy}}, \bibinfo {author} {\bibfnamefont {A.}~\bibnamefont
  {Obertelli}}, \bibinfo {author} {\bibfnamefont {S.}~\bibnamefont {P\'eru}},
  \bibinfo {author} {\bibfnamefont {N.}~\bibnamefont {Pillet}}, \bibinfo
  {author} {\bibfnamefont {S.}~\bibnamefont {Hilaire}}, \bibinfo {author}
  {\bibfnamefont {J.~P.}\ \bibnamefont {Delaroche}}, \bibinfo {author}
  {\bibfnamefont {M.}~\bibnamefont {Girod}}, \ and\ \bibinfo {author}
  {\bibfnamefont {J.}~\bibnamefont {Libert}},\ }\href {\doibase
  10.1103/PhysRevC.80.064313} {\bibfield  {journal} {\bibinfo  {journal} {Phys.
  Rev. C}\ }\textbf {\bibinfo {volume} {80}},\ \bibinfo {pages} {064313}
  (\bibinfo {year} {2009})}\BibitemShut {NoStop}%
\bibitem [{\citenamefont {Nik\ifmmode \check{s}\else
  \v{s}\fi{}i\ifmmode~\acute{c}\else \'{c}\fi{}}\ \emph
  {et~al.}(2014)\citenamefont {Nik\ifmmode \check{s}\else
  \v{s}\fi{}i\ifmmode~\acute{c}\else \'{c}\fi{}}, \citenamefont
  {Marevi\ifmmode~\acute{c}\else \'{c}\fi{}},\ and\ \citenamefont
  {Vretenar}}]{niksic2014}%
  \BibitemOpen
  \bibfield  {author} {\bibinfo {author} {\bibfnamefont {T.}~\bibnamefont
  {Nik\ifmmode \check{s}\else \v{s}\fi{}i\ifmmode~\acute{c}\else \'{c}\fi{}}},
  \bibinfo {author} {\bibfnamefont {P.}~\bibnamefont
  {Marevi\ifmmode~\acute{c}\else \'{c}\fi{}}}, \ and\ \bibinfo {author}
  {\bibfnamefont {D.}~\bibnamefont {Vretenar}},\ }\href {\doibase
  10.1103/PhysRevC.89.044325} {\bibfield  {journal} {\bibinfo  {journal} {Phys.
  Rev. C}\ }\textbf {\bibinfo {volume} {89}},\ \bibinfo {pages} {044325}
  (\bibinfo {year} {2014})}\BibitemShut {NoStop}%
\bibitem [{\citenamefont {Wang}\ \emph {et~al.}(2015)\citenamefont {Wang},
  \citenamefont {Xiang}, \citenamefont {Long},\ and\ \citenamefont
  {Li}}]{wang2015}%
  \BibitemOpen
  \bibfield  {author} {\bibinfo {author} {\bibfnamefont {Z.~H.}\ \bibnamefont
  {Wang}}, \bibinfo {author} {\bibfnamefont {J.}~\bibnamefont {Xiang}},
  \bibinfo {author} {\bibfnamefont {W.~H.}\ \bibnamefont {Long}}, \ and\
  \bibinfo {author} {\bibfnamefont {Z.~P.}\ \bibnamefont {Li}},\ }\href
  {http://stacks.iop.org/0954-3899/42/i=4/a=045108} {\bibfield  {journal}
  {\bibinfo  {journal} {Journal of Physics G: Nuclear and Particle Physics}\
  }\textbf {\bibinfo {volume} {42}},\ \bibinfo {pages} {045108} (\bibinfo
  {year} {2015})}\BibitemShut {NoStop}%
\bibitem [{\citenamefont {Sarriguren}(2015)}]{sarriguren2015}%
  \BibitemOpen
  \bibfield  {author} {\bibinfo {author} {\bibfnamefont {P.}~\bibnamefont
  {Sarriguren}},\ }\href@noop {} {\bibfield  {journal} {\bibinfo  {journal}
  {Phys. Rev. C}\ }\textbf {\bibinfo {volume} {91}},\ \bibinfo {pages} {044304}
  (\bibinfo {year} {2015})}\BibitemShut {NoStop}%
\bibitem [{\citenamefont {Padilla-Rodal}\ \emph {et~al.}(2006)\citenamefont
  {Padilla-Rodal}, \citenamefont {Castanos}, \citenamefont {Bijker},\ and\
  \citenamefont {Galindo-Uribarri}}]{padilla2006}%
  \BibitemOpen
  \bibfield  {author} {\bibinfo {author} {\bibfnamefont {E.}~\bibnamefont
  {Padilla-Rodal}}, \bibinfo {author} {\bibfnamefont {O.}~\bibnamefont
  {Castanos}}, \bibinfo {author} {\bibfnamefont {R.}~\bibnamefont {Bijker}}, \
  and\ \bibinfo {author} {\bibfnamefont {A.}~\bibnamefont {Galindo-Uribarri}},\
  }\href@noop {} {\bibfield  {journal} {\bibinfo  {journal} {Rev. Mex. Fis. S}\
  }\textbf {\bibinfo {volume} {52}},\ \bibinfo {pages} {57} (\bibinfo {year}
  {2006})}\BibitemShut {NoStop}%
\bibitem [{\citenamefont {Barea}\ and\ \citenamefont
  {Iachello}(2009)}]{barea2009}%
  \BibitemOpen
  \bibfield  {author} {\bibinfo {author} {\bibfnamefont {J.}~\bibnamefont
  {Barea}}\ and\ \bibinfo {author} {\bibfnamefont {F.}~\bibnamefont
  {Iachello}},\ }\href {\doibase 10.1103/PhysRevC.79.044301} {\bibfield
  {journal} {\bibinfo  {journal} {Phys. Rev. C}\ }\textbf {\bibinfo {volume}
  {79}},\ \bibinfo {pages} {044301} (\bibinfo {year} {2009})}\BibitemShut
  {NoStop}%
\bibitem [{\citenamefont {Heyde}\ and\ \citenamefont {Wood}(2011)}]{heyde2011}%
  \BibitemOpen
  \bibfield  {author} {\bibinfo {author} {\bibfnamefont {K.}~\bibnamefont
  {Heyde}}\ and\ \bibinfo {author} {\bibfnamefont {J.~L.}\ \bibnamefont
  {Wood}},\ }\href {\doibase 10.1103/RevModPhys.83.1467} {\bibfield  {journal}
  {\bibinfo  {journal} {Rev. Mod. Phys.}\ }\textbf {\bibinfo {volume} {83}},\
  \bibinfo {pages} {1467} (\bibinfo {year} {2011})}\BibitemShut {NoStop}%
\bibitem [{\citenamefont {Bender}\ \emph {et~al.}(2003)\citenamefont {Bender},
  \citenamefont {Heenen},\ and\ \citenamefont {Reinhard}}]{bender2003}%
  \BibitemOpen
  \bibfield  {author} {\bibinfo {author} {\bibfnamefont {M.}~\bibnamefont
  {Bender}}, \bibinfo {author} {\bibfnamefont {P.-H.}\ \bibnamefont {Heenen}},
  \ and\ \bibinfo {author} {\bibfnamefont {P.-G.}\ \bibnamefont {Reinhard}},\
  }\href {\doibase 10.1103/RevModPhys.75.121} {\bibfield  {journal} {\bibinfo
  {journal} {Rev. Mod. Phys.}\ }\textbf {\bibinfo {volume} {75}},\ \bibinfo
  {pages} {121} (\bibinfo {year} {2003})}\BibitemShut {NoStop}%
\bibitem [{\citenamefont {Skyrme}(1958)}]{Skyrme}%
  \BibitemOpen
  \bibfield  {author} {\bibinfo {author} {\bibfnamefont {T.~H.~R.}\
  \bibnamefont {Skyrme}},\ }\href@noop {} {\bibfield  {journal} {\bibinfo
  {journal} {Nucl. Phys.}\ }\textbf {\bibinfo {volume} {9}},\ \bibinfo {pages}
  {615} (\bibinfo {year} {1958})}\BibitemShut {NoStop}%
\bibitem [{\citenamefont {{J. Decharge and M. Girod and D.
  Gogny}}(1975)}]{Gogny}%
  \BibitemOpen
  \bibfield  {author} {\bibinfo {author} {\bibnamefont {{J. Decharge and M.
  Girod and D. Gogny}}},\ }\href {\doibase 10.1016/0370-2693(75)90359-7}
  {\bibfield  {journal} {\bibinfo  {journal} {Phys. Lett. B}\ }\textbf
  {\bibinfo {volume} {55}},\ \bibinfo {pages} {361} (\bibinfo {year}
  {1975})}\BibitemShut {NoStop}%
\bibitem [{\citenamefont {Vretenar}\ \emph {et~al.}(2005)\citenamefont
  {Vretenar}, \citenamefont {Afanasjev}, \citenamefont {Lalazissis},\ and\
  \citenamefont {Ring}}]{vretenar2005}%
  \BibitemOpen
  \bibfield  {author} {\bibinfo {author} {\bibfnamefont {D.}~\bibnamefont
  {Vretenar}}, \bibinfo {author} {\bibfnamefont {A.~V.}\ \bibnamefont
  {Afanasjev}}, \bibinfo {author} {\bibfnamefont {G.}~\bibnamefont
  {Lalazissis}}, \ and\ \bibinfo {author} {\bibfnamefont {P.}~\bibnamefont
  {Ring}},\ }\href {\doibase 10.1016/j.physrep.2004.10.001} {\bibfield
  {journal} {\bibinfo  {journal} {Phys. Rep.}\ }\textbf {\bibinfo {volume}
  {409}},\ \bibinfo {pages} {101} (\bibinfo {year} {2005})}\BibitemShut
  {NoStop}%
\bibitem [{\citenamefont {Nik\ifmmode \check{s}\else
  \v{s}\fi{}i\ifmmode~\acute{c}\else \'{c}\fi{}}\ \emph
  {et~al.}(2011)\citenamefont {Nik\ifmmode \check{s}\else
  \v{s}\fi{}i\ifmmode~\acute{c}\else \'{c}\fi{}}, \citenamefont {Vretenar},\
  and\ \citenamefont {Ring}}]{niksic2011}%
  \BibitemOpen
  \bibfield  {author} {\bibinfo {author} {\bibfnamefont {T.}~\bibnamefont
  {Nik\ifmmode \check{s}\else \v{s}\fi{}i\ifmmode~\acute{c}\else \'{c}\fi{}}},
  \bibinfo {author} {\bibfnamefont {D.}~\bibnamefont {Vretenar}}, \ and\
  \bibinfo {author} {\bibfnamefont {P.}~\bibnamefont {Ring}},\ }\href {\doibase
  10.1016/j.ppnp.2011.01.055} {\bibfield  {journal} {\bibinfo  {journal} {Prog.
  Part. Nucl. Phys.}\ }\textbf {\bibinfo {volume} {66}},\ \bibinfo {pages}
  {519} (\bibinfo {year} {2011})}\BibitemShut {NoStop}%
\bibitem [{\citenamefont {Mei}\ \emph {et~al.}(2012)\citenamefont {Mei},
  \citenamefont {Xiang}, \citenamefont {Yao}, \citenamefont {Li},\ and\
  \citenamefont {Meng}}]{mei2012}%
  \BibitemOpen
  \bibfield  {author} {\bibinfo {author} {\bibfnamefont {H.}~\bibnamefont
  {Mei}}, \bibinfo {author} {\bibfnamefont {J.}~\bibnamefont {Xiang}}, \bibinfo
  {author} {\bibfnamefont {J.~M.}\ \bibnamefont {Yao}}, \bibinfo {author}
  {\bibfnamefont {Z.~P.}\ \bibnamefont {Li}}, \ and\ \bibinfo {author}
  {\bibfnamefont {J.}~\bibnamefont {Meng}},\ }\href@noop {} {\bibfield
  {journal} {\bibinfo  {journal} {Phys. Rev. C}\ }\textbf {\bibinfo {volume}
  {85}},\ \bibinfo {pages} {034321} (\bibinfo {year} {2012})}\BibitemShut
  {NoStop}%
\bibitem [{\citenamefont {Rodr\'{\i}guez}(2014)}]{trodriguez2014}%
  \BibitemOpen
  \bibfield  {author} {\bibinfo {author} {\bibfnamefont {T.~R.}\ \bibnamefont
  {Rodr\'{\i}guez}},\ }\href@noop {} {\bibfield  {journal} {\bibinfo  {journal}
  {Phys. Rev. C}\ }\textbf {\bibinfo {volume} {90}},\ \bibinfo {pages} {034306}
  (\bibinfo {year} {2014})}\BibitemShut {NoStop}%
\bibitem [{\citenamefont {Ring}\ and\ \citenamefont {Schuck}(1980)}]{RS}%
  \BibitemOpen
  \bibfield  {author} {\bibinfo {author} {\bibfnamefont {P.}~\bibnamefont
  {Ring}}\ and\ \bibinfo {author} {\bibfnamefont {P.}~\bibnamefont {Schuck}},\
  }\href@noop {} {\emph {\bibinfo {title} {The nuclear many-body problem}}}\
  (\bibinfo  {publisher} {Berlin: Springer-Verlag},\ \bibinfo {year}
  {1980})\BibitemShut {NoStop}%
\bibitem [{\citenamefont {Rodr\'iguez-Guzm\'an}\ \emph
  {et~al.}(2002)\citenamefont {Rodr\'iguez-Guzm\'an}, \citenamefont {Egido},\
  and\ \citenamefont {Robledo}}]{rayner2002}%
  \BibitemOpen
  \bibfield  {author} {\bibinfo {author} {\bibfnamefont {R.}~\bibnamefont
  {Rodr\'iguez-Guzm\'an}}, \bibinfo {author} {\bibfnamefont {J.~L.}\
  \bibnamefont {Egido}}, \ and\ \bibinfo {author} {\bibfnamefont {L.~M.}\
  \bibnamefont {Robledo}},\ }\href@noop {} {\bibfield  {journal} {\bibinfo
  {journal} {Nucl. Phys. A}\ }\textbf {\bibinfo {volume} {709}},\ \bibinfo
  {pages} {201 } (\bibinfo {year} {2002})}\BibitemShut {NoStop}%
\bibitem [{\citenamefont {Li}\ \emph {et~al.}(2016)\citenamefont {Li},
  \citenamefont {Nik\ifmmode \check{s}\else \v{s}\fi{}i\ifmmode~\acute{c}\else
  \'{c}\fi{}},\ and\ \citenamefont {Vretenar}}]{li2016}%
  \BibitemOpen
  \bibfield  {author} {\bibinfo {author} {\bibfnamefont {Z.~P.}\ \bibnamefont
  {Li}}, \bibinfo {author} {\bibfnamefont {T.}~\bibnamefont {Nik\ifmmode
  \check{s}\else \v{s}\fi{}i\ifmmode~\acute{c}\else \'{c}\fi{}}}, \ and\
  \bibinfo {author} {\bibfnamefont {D.}~\bibnamefont {Vretenar}},\ }\href@noop
  {} {\bibfield  {journal} {\bibinfo  {journal} {J. Phys. G: Nucl. Part.
  Phys.}\ }\textbf {\bibinfo {volume} {43}},\ \bibinfo {pages} {024005}
  (\bibinfo {year} {2016})}\BibitemShut {NoStop}%
\bibitem [{\citenamefont {Petrovici}\ \emph {et~al.}(1988)\citenamefont
  {Petrovici}, \citenamefont {Schmid}, \citenamefont {Gr^^c3^^bcmmer},
  \citenamefont {Faessler},\ and\ \citenamefont {Horibata}}]{Petrovici1988317}%
  \BibitemOpen
  \bibfield  {author} {\bibinfo {author} {\bibfnamefont {A.}~\bibnamefont
  {Petrovici}}, \bibinfo {author} {\bibfnamefont {K.}~\bibnamefont {Schmid}},
  \bibinfo {author} {\bibfnamefont {F.}~\bibnamefont {Gr^^c3^^bcmmer}},
  \bibinfo {author} {\bibfnamefont {A.}~\bibnamefont {Faessler}}, \ and\
  \bibinfo {author} {\bibfnamefont {T.}~\bibnamefont {Horibata}},\ }\href
  {\doibase http://dx.doi.org/10.1016/0375-9474(88)90539-8} {\bibfield
  {journal} {\bibinfo  {journal} {Nuclear Physics A}\ }\textbf {\bibinfo
  {volume} {483}},\ \bibinfo {pages} {317 } (\bibinfo {year}
  {1988})}\BibitemShut {NoStop}%
\bibitem [{\citenamefont {Petrovici}\ \emph {et~al.}(1989)\citenamefont
  {Petrovici}, \citenamefont {Schmid}, \citenamefont {Gr^^c3^^bcmmer},\ and\
  \citenamefont {Faessler}}]{Petrovici1989277}%
  \BibitemOpen
  \bibfield  {author} {\bibinfo {author} {\bibfnamefont {A.}~\bibnamefont
  {Petrovici}}, \bibinfo {author} {\bibfnamefont {K.}~\bibnamefont {Schmid}},
  \bibinfo {author} {\bibfnamefont {F.}~\bibnamefont {Gr^^c3^^bcmmer}}, \ and\
  \bibinfo {author} {\bibfnamefont {A.}~\bibnamefont {Faessler}},\ }\href
  {\doibase https://doi.org/10.1016/0375-9474(89)90346-1} {\bibfield  {journal}
  {\bibinfo  {journal} {Nuclear Physics A}\ }\textbf {\bibinfo {volume}
  {504}},\ \bibinfo {pages} {277 } (\bibinfo {year} {1989})}\BibitemShut
  {NoStop}%
\bibitem [{\citenamefont {Petrovici}\ \emph {et~al.}(1990)\citenamefont
  {Petrovici}, \citenamefont {Schmid}, \citenamefont {Gr^^c3^^bcmmer},\ and\
  \citenamefont {Faessler}}]{Petrovici1990108}%
  \BibitemOpen
  \bibfield  {author} {\bibinfo {author} {\bibfnamefont {A.}~\bibnamefont
  {Petrovici}}, \bibinfo {author} {\bibfnamefont {K.}~\bibnamefont {Schmid}},
  \bibinfo {author} {\bibfnamefont {F.}~\bibnamefont {Gr^^c3^^bcmmer}}, \ and\
  \bibinfo {author} {\bibfnamefont {A.}~\bibnamefont {Faessler}},\ }\href
  {\doibase https://doi.org/10.1016/0375-9474(90)90263-L} {\bibfield  {journal}
  {\bibinfo  {journal} {Nuclear Physics A}\ }\textbf {\bibinfo {volume}
  {517}},\ \bibinfo {pages} {108 } (\bibinfo {year} {1990})}\BibitemShut
  {NoStop}%
\bibitem [{\citenamefont {Petrovici}\ \emph {et~al.}(1992)\citenamefont
  {Petrovici}, \citenamefont {Hammar^^c3^^a9n}, \citenamefont {Schmid},
  \citenamefont {Gr^^c3^^bcmmer},\ and\ \citenamefont
  {Faessler}}]{PETROVICI1992352}%
  \BibitemOpen
  \bibfield  {author} {\bibinfo {author} {\bibfnamefont {A.}~\bibnamefont
  {Petrovici}}, \bibinfo {author} {\bibfnamefont {E.}~\bibnamefont
  {Hammar^^c3^^a9n}}, \bibinfo {author} {\bibfnamefont {K.}~\bibnamefont
  {Schmid}}, \bibinfo {author} {\bibfnamefont {F.}~\bibnamefont
  {Gr^^c3^^bcmmer}}, \ and\ \bibinfo {author} {\bibfnamefont {A.}~\bibnamefont
  {Faessler}},\ }\href {\doibase
  http://dx.doi.org/10.1016/0375-9474(92)90084-W} {\bibfield  {journal}
  {\bibinfo  {journal} {Nuclear Physics A}\ }\textbf {\bibinfo {volume}
  {549}},\ \bibinfo {pages} {352 } (\bibinfo {year} {1992})}\BibitemShut
  {NoStop}%
\bibitem [{\citenamefont {Petrovici}\ \emph {et~al.}(2002)\citenamefont
  {Petrovici}, \citenamefont {Schmid},\ and\ \citenamefont
  {Faessler}}]{Petrovici2002246}%
  \BibitemOpen
  \bibfield  {author} {\bibinfo {author} {\bibfnamefont {A.}~\bibnamefont
  {Petrovici}}, \bibinfo {author} {\bibfnamefont {K.}~\bibnamefont {Schmid}}, \
  and\ \bibinfo {author} {\bibfnamefont {A.}~\bibnamefont {Faessler}},\ }\href
  {\doibase https://doi.org/10.1016/S0375-9474(02)01089-8} {\bibfield
  {journal} {\bibinfo  {journal} {Nuclear Physics A}\ }\textbf {\bibinfo
  {volume} {710}},\ \bibinfo {pages} {246 } (\bibinfo {year}
  {2002})}\BibitemShut {NoStop}%
\bibitem [{\citenamefont {Nomura}\ \emph {et~al.}(2008)\citenamefont {Nomura},
  \citenamefont {Shimizu},\ and\ \citenamefont {Otsuka}}]{nomura2008}%
  \BibitemOpen
  \bibfield  {author} {\bibinfo {author} {\bibfnamefont {K.}~\bibnamefont
  {Nomura}}, \bibinfo {author} {\bibfnamefont {N.}~\bibnamefont {Shimizu}}, \
  and\ \bibinfo {author} {\bibfnamefont {T.}~\bibnamefont {Otsuka}},\ }\href
  {\doibase 10.1103/PhysRevLett.101.142501} {\bibfield  {journal} {\bibinfo
  {journal} {Phys. Rev. Lett.}\ }\textbf {\bibinfo {volume} {101}},\ \bibinfo
  {pages} {142501} (\bibinfo {year} {2008})}\BibitemShut {NoStop}%
\bibitem [{\citenamefont {Iachello}\ and\ \citenamefont {Arima}(1987)}]{IBM}%
  \BibitemOpen
  \bibfield  {author} {\bibinfo {author} {\bibfnamefont {F.}~\bibnamefont
  {Iachello}}\ and\ \bibinfo {author} {\bibfnamefont {A.}~\bibnamefont
  {Arima}},\ }\href@noop {} {\emph {\bibinfo {title} {The interacting boson
  model}}}\ (\bibinfo  {publisher} {Cambridge University Press, Cambridge},\
  \bibinfo {year} {1987})\BibitemShut {NoStop}%
\bibitem [{\citenamefont {Nomura}\ \emph
  {et~al.}(2016{\natexlab{a}})\citenamefont {Nomura}, \citenamefont {Otsuka},\
  and\ \citenamefont {{Van Isacker}}}]{nomura2016}%
  \BibitemOpen
  \bibfield  {author} {\bibinfo {author} {\bibfnamefont {K.}~\bibnamefont
  {Nomura}}, \bibinfo {author} {\bibfnamefont {T.}~\bibnamefont {Otsuka}}, \
  and\ \bibinfo {author} {\bibfnamefont {P.}~\bibnamefont {{Van Isacker}}},\
  }\href@noop {} {\bibfield  {journal} {\bibinfo  {journal} {J. Phys. G: Nucl.
  Part. Phys.}\ }\textbf {\bibinfo {volume} {43}},\ \bibinfo {pages} {024008}
  (\bibinfo {year} {2016}{\natexlab{a}})}\BibitemShut {NoStop}%
\bibitem [{\citenamefont {Nomura}\ \emph
  {et~al.}(2016{\natexlab{b}})\citenamefont {Nomura}, \citenamefont
  {Rodr\'{\i}guez-Guzm\'an},\ and\ \citenamefont {Robledo}}]{nomura2016zr}%
  \BibitemOpen
  \bibfield  {author} {\bibinfo {author} {\bibfnamefont {K.}~\bibnamefont
  {Nomura}}, \bibinfo {author} {\bibfnamefont {R.}~\bibnamefont
  {Rodr\'{\i}guez-Guzm\'an}}, \ and\ \bibinfo {author} {\bibfnamefont {L.~M.}\
  \bibnamefont {Robledo}},\ }\href {\doibase 10.1103/PhysRevC.94.044314}
  {\bibfield  {journal} {\bibinfo  {journal} {Phys. Rev. C}\ }\textbf {\bibinfo
  {volume} {94}},\ \bibinfo {pages} {044314} (\bibinfo {year}
  {2016}{\natexlab{b}})}\BibitemShut {NoStop}%
\bibitem [{\citenamefont {Nomura}\ \emph {et~al.}(2014)\citenamefont {Nomura},
  \citenamefont {Vretenar}, \citenamefont {Nik\ifmmode \check{s}\else
  \v{s}\fi{}i\ifmmode~\acute{c}\else \'{c}\fi{}},\ and\ \citenamefont
  {Lu}}]{nomura2014}%
  \BibitemOpen
  \bibfield  {author} {\bibinfo {author} {\bibfnamefont {K.}~\bibnamefont
  {Nomura}}, \bibinfo {author} {\bibfnamefont {D.}~\bibnamefont {Vretenar}},
  \bibinfo {author} {\bibfnamefont {T.}~\bibnamefont {Nik\ifmmode
  \check{s}\else \v{s}\fi{}i\ifmmode~\acute{c}\else \'{c}\fi{}}}, \ and\
  \bibinfo {author} {\bibfnamefont {B.-N.}\ \bibnamefont {Lu}},\ }\href
  {\doibase 10.1103/PhysRevC.89.024312} {\bibfield  {journal} {\bibinfo
  {journal} {Phys. Rev. C}\ }\textbf {\bibinfo {volume} {89}},\ \bibinfo
  {pages} {024312} (\bibinfo {year} {2014})}\BibitemShut {NoStop}%
\bibitem [{\citenamefont {Nomura}\ \emph {et~al.}(2015)\citenamefont {Nomura},
  \citenamefont {Rodr\'{\i}guez-Guzm\'an},\ and\ \citenamefont
  {Robledo}}]{nomura2015}%
  \BibitemOpen
  \bibfield  {author} {\bibinfo {author} {\bibfnamefont {K.}~\bibnamefont
  {Nomura}}, \bibinfo {author} {\bibfnamefont {R.}~\bibnamefont
  {Rodr\'{\i}guez-Guzm\'an}}, \ and\ \bibinfo {author} {\bibfnamefont {L.~M.}\
  \bibnamefont {Robledo}},\ }\href@noop {} {\bibfield  {journal} {\bibinfo
  {journal} {Phys. Rev. C}\ }\textbf {\bibinfo {volume} {92}},\ \bibinfo
  {pages} {014312} (\bibinfo {year} {2015})}\BibitemShut {NoStop}%
\bibitem [{\citenamefont {Nomura}\ \emph
  {et~al.}(2016{\natexlab{c}})\citenamefont {Nomura}, \citenamefont
  {Nik\ifmmode \check{s}\else \v{s}\fi{}i\ifmmode~\acute{c}\else \'{c}\fi{}},\
  and\ \citenamefont {Vretenar}}]{nomura2016odd}%
  \BibitemOpen
  \bibfield  {author} {\bibinfo {author} {\bibfnamefont {K.}~\bibnamefont
  {Nomura}}, \bibinfo {author} {\bibfnamefont {T.}~\bibnamefont {Nik\ifmmode
  \check{s}\else \v{s}\fi{}i\ifmmode~\acute{c}\else \'{c}\fi{}}}, \ and\
  \bibinfo {author} {\bibfnamefont {D.}~\bibnamefont {Vretenar}},\ }\href
  {\doibase 10.1103/PhysRevC.93.054305} {\bibfield  {journal} {\bibinfo
  {journal} {Phys. Rev. C}\ }\textbf {\bibinfo {volume} {93}},\ \bibinfo
  {pages} {054305} (\bibinfo {year} {2016}{\natexlab{c}})}\BibitemShut
  {NoStop}%
\bibitem [{\citenamefont {Duval}\ \emph {et~al.}(1983)\citenamefont {Duval},
  \citenamefont {Goutte},\ and\ \citenamefont {Vergnes}}]{duval1983}%
  \BibitemOpen
  \bibfield  {author} {\bibinfo {author} {\bibfnamefont {P.}~\bibnamefont
  {Duval}}, \bibinfo {author} {\bibfnamefont {D.}~\bibnamefont {Goutte}}, \
  and\ \bibinfo {author} {\bibfnamefont {M.}~\bibnamefont {Vergnes}},\ }\href
  {\doibase http://dx.doi.org/10.1016/0370-2693(83)91457-0} {\bibfield
  {journal} {\bibinfo  {journal} {Physics Letters B}\ }\textbf {\bibinfo
  {volume} {124}},\ \bibinfo {pages} {297 } (\bibinfo {year}
  {1983})}\BibitemShut {NoStop}%
\bibitem [{\citenamefont {Kaup}\ \emph {et~al.}(1983)\citenamefont {Kaup},
  \citenamefont {M{\"o}nkemeyer},\ and\ \citenamefont {Brentano}}]{kaup1983}%
  \BibitemOpen
  \bibfield  {author} {\bibinfo {author} {\bibfnamefont {U.}~\bibnamefont
  {Kaup}}, \bibinfo {author} {\bibfnamefont {C.}~\bibnamefont
  {M{\"o}nkemeyer}}, \ and\ \bibinfo {author} {\bibfnamefont {P.~v.}\
  \bibnamefont {Brentano}},\ }\href {\doibase 10.1007/BF01433622} {\bibfield
  {journal} {\bibinfo  {journal} {Zeitschrift f{\"u}r Physik A Atoms and
  Nuclei}\ }\textbf {\bibinfo {volume} {310}},\ \bibinfo {pages} {129}
  (\bibinfo {year} {1983})}\BibitemShut {NoStop}%
\bibitem [{\citenamefont {Goriely}\ \emph {et~al.}(2009)\citenamefont
  {Goriely}, \citenamefont {Hilaire}, \citenamefont {Girod},\ and\
  \citenamefont {P\'eru}}]{D1M}%
  \BibitemOpen
  \bibfield  {author} {\bibinfo {author} {\bibfnamefont {S.}~\bibnamefont
  {Goriely}}, \bibinfo {author} {\bibfnamefont {S.}~\bibnamefont {Hilaire}},
  \bibinfo {author} {\bibfnamefont {M.}~\bibnamefont {Girod}}, \ and\ \bibinfo
  {author} {\bibfnamefont {S.}~\bibnamefont {P\'eru}},\ }\href {\doibase
  10.1103/PhysRevLett.102.242501} {\bibfield  {journal} {\bibinfo  {journal}
  {Phys. Rev. Lett.}\ }\textbf {\bibinfo {volume} {102}},\ \bibinfo {pages}
  {242501} (\bibinfo {year} {2009})}\BibitemShut {NoStop}%
\bibitem [{\citenamefont {Berger}\ \emph {et~al.}(1984)\citenamefont {Berger},
  \citenamefont {Girod},\ and\ \citenamefont {Gogny}}]{D1S}%
  \BibitemOpen
  \bibfield  {author} {\bibinfo {author} {\bibfnamefont {J.~F.}\ \bibnamefont
  {Berger}}, \bibinfo {author} {\bibfnamefont {M.}~\bibnamefont {Girod}}, \
  and\ \bibinfo {author} {\bibfnamefont {D.}~\bibnamefont {Gogny}},\
  }\href@noop {} {\bibfield  {journal} {\bibinfo  {journal} {Nucl. Phys. A}\
  }\textbf {\bibinfo {volume} {428}},\ \bibinfo {pages} {23 } (\bibinfo {year}
  {1984})}\BibitemShut {NoStop}%
\bibitem [{\citenamefont {Robledo}\ \emph {et~al.}(2008)\citenamefont
  {Robledo}, \citenamefont {Rodr\'{\i}guez-Guzm\'an},\ and\ \citenamefont
  {Sarriguren}}]{robledo2008}%
  \BibitemOpen
  \bibfield  {author} {\bibinfo {author} {\bibfnamefont {L.~M.}\ \bibnamefont
  {Robledo}}, \bibinfo {author} {\bibfnamefont {R.~R.}\ \bibnamefont
  {Rodr\'{\i}guez-Guzm\'an}}, \ and\ \bibinfo {author} {\bibfnamefont
  {P.}~\bibnamefont {Sarriguren}},\ }\href@noop {} {\bibfield  {journal}
  {\bibinfo  {journal} {Phys. Rev. C}\ }\textbf {\bibinfo {volume} {78}},\
  \bibinfo {pages} {034314} (\bibinfo {year} {2008})}\BibitemShut {NoStop}%
\bibitem [{\citenamefont {Rodr\'iguez-Guzm\'an}\ \emph
  {et~al.}(2010)\citenamefont {Rodr\'iguez-Guzm\'an}, \citenamefont
  {Sarriguren}, \citenamefont {Robledo},\ and\ \citenamefont
  {Garc\'ia-Ramos}}]{rayner2010pt}%
  \BibitemOpen
  \bibfield  {author} {\bibinfo {author} {\bibfnamefont {R.}~\bibnamefont
  {Rodr\'iguez-Guzm\'an}}, \bibinfo {author} {\bibfnamefont {P.}~\bibnamefont
  {Sarriguren}}, \bibinfo {author} {\bibfnamefont {L.~M.}\ \bibnamefont
  {Robledo}}, \ and\ \bibinfo {author} {\bibfnamefont {J.~E.}\ \bibnamefont
  {Garc\'ia-Ramos}},\ }\href {\doibase 10.1103/PhysRevC.81.024310} {\bibfield
  {journal} {\bibinfo  {journal} {Phys. Rev. C}\ }\textbf {\bibinfo {volume}
  {81}},\ \bibinfo {pages} {024310} (\bibinfo {year} {2010})}\BibitemShut
  {NoStop}%
\bibitem [{\citenamefont {Ginocchio}\ and\ \citenamefont {Kirson}(1980)}]{GK}%
  \BibitemOpen
  \bibfield  {author} {\bibinfo {author} {\bibfnamefont {J.~N.}\ \bibnamefont
  {Ginocchio}}\ and\ \bibinfo {author} {\bibfnamefont {M.~W.}\ \bibnamefont
  {Kirson}},\ }\href {\doibase 10.1016/0375-9474(80)90387-5} {\bibfield
  {journal} {\bibinfo  {journal} {Nucl. Phys. A}\ }\textbf {\bibinfo {volume}
  {350}},\ \bibinfo {pages} {31} (\bibinfo {year} {1980})}\BibitemShut
  {NoStop}%
\bibitem [{\citenamefont {Otsuka}\ \emph {et~al.}(1978)\citenamefont {Otsuka},
  \citenamefont {Arima},\ and\ \citenamefont {Iachello}}]{OAI}%
  \BibitemOpen
  \bibfield  {author} {\bibinfo {author} {\bibfnamefont {T.}~\bibnamefont
  {Otsuka}}, \bibinfo {author} {\bibfnamefont {A.}~\bibnamefont {Arima}}, \
  and\ \bibinfo {author} {\bibfnamefont {F.}~\bibnamefont {Iachello}},\
  }\href@noop {} {\bibfield  {journal} {\bibinfo  {journal} {Nucl. Phys. A}\
  }\textbf {\bibinfo {volume} {309}},\ \bibinfo {pages} {1} (\bibinfo {year}
  {1978})}\BibitemShut {NoStop}%
\bibitem [{\citenamefont {Duval}\ and\ \citenamefont
  {Barrett}(1981)}]{duval81}%
  \BibitemOpen
  \bibfield  {author} {\bibinfo {author} {\bibfnamefont {P.~D.}\ \bibnamefont
  {Duval}}\ and\ \bibinfo {author} {\bibfnamefont {B.~R.}\ \bibnamefont
  {Barrett}},\ }\href@noop {} {\bibfield  {journal} {\bibinfo  {journal} {Phys.
  Lett. B}\ }\textbf {\bibinfo {volume} {100}},\ \bibinfo {pages} {223}
  (\bibinfo {year} {1981})}\BibitemShut {NoStop}%
\bibitem [{\citenamefont {Nomura}\ \emph
  {et~al.}(2012{\natexlab{a}})\citenamefont {Nomura}, \citenamefont {Shimizu},
  \citenamefont {Vretenar}, \citenamefont {Nik\ifmmode \check{s}\else
  \v{s}\fi{}i\ifmmode~\acute{c}\else \'{c}\fi{}},\ and\ \citenamefont
  {Otsuka}}]{nomura2012tri}%
  \BibitemOpen
  \bibfield  {author} {\bibinfo {author} {\bibfnamefont {K.}~\bibnamefont
  {Nomura}}, \bibinfo {author} {\bibfnamefont {N.}~\bibnamefont {Shimizu}},
  \bibinfo {author} {\bibfnamefont {D.}~\bibnamefont {Vretenar}}, \bibinfo
  {author} {\bibfnamefont {T.}~\bibnamefont {Nik\ifmmode \check{s}\else
  \v{s}\fi{}i\ifmmode~\acute{c}\else \'{c}\fi{}}}, \ and\ \bibinfo {author}
  {\bibfnamefont {T.}~\bibnamefont {Otsuka}},\ }\href {\doibase
  10.1103/PhysRevLett.108.132501} {\bibfield  {journal} {\bibinfo  {journal}
  {Phys. Rev. Lett.}\ }\textbf {\bibinfo {volume} {108}},\ \bibinfo {pages}
  {132501} (\bibinfo {year} {2012}{\natexlab{a}})}\BibitemShut {NoStop}%
\bibitem [{\citenamefont {Frank}\ \emph {et~al.}(2004)\citenamefont {Frank},
  \citenamefont {Van~Isacker},\ and\ \citenamefont {Vargas}}]{frank04}%
  \BibitemOpen
  \bibfield  {author} {\bibinfo {author} {\bibfnamefont {A.}~\bibnamefont
  {Frank}}, \bibinfo {author} {\bibfnamefont {P.}~\bibnamefont {Van~Isacker}},
  \ and\ \bibinfo {author} {\bibfnamefont {C.~E.}\ \bibnamefont {Vargas}},\
  }\href {\doibase 10.1103/PhysRevC.69.034323} {\bibfield  {journal} {\bibinfo
  {journal} {Phys. Rev. C}\ }\textbf {\bibinfo {volume} {69}},\ \bibinfo
  {pages} {034323} (\bibinfo {year} {2004})}\BibitemShut {NoStop}%
\bibitem [{\citenamefont {Nomura}\ \emph
  {et~al.}(2012{\natexlab{b}})\citenamefont {Nomura}, \citenamefont
  {Rodr\'iguez-Guzm\'an}, \citenamefont {Robledo},\ and\ \citenamefont
  {Shimizu}}]{nomura2012sc}%
  \BibitemOpen
  \bibfield  {author} {\bibinfo {author} {\bibfnamefont {K.}~\bibnamefont
  {Nomura}}, \bibinfo {author} {\bibfnamefont {R.}~\bibnamefont
  {Rodr\'iguez-Guzm\'an}}, \bibinfo {author} {\bibfnamefont {L.~M.}\
  \bibnamefont {Robledo}}, \ and\ \bibinfo {author} {\bibfnamefont
  {N.}~\bibnamefont {Shimizu}},\ }\href {\doibase 10.1103/PhysRevC.86.034322}
  {\bibfield  {journal} {\bibinfo  {journal} {Phys. Rev. C}\ }\textbf {\bibinfo
  {volume} {86}},\ \bibinfo {pages} {034322} (\bibinfo {year}
  {2012}{\natexlab{b}})}\BibitemShut {NoStop}%
\bibitem [{\citenamefont {Nomura}\ \emph {et~al.}(2013)\citenamefont {Nomura},
  \citenamefont {Rodr\'{\i}guez-Guzm\'an},\ and\ \citenamefont
  {Robledo}}]{nomura2013hg}%
  \BibitemOpen
  \bibfield  {author} {\bibinfo {author} {\bibfnamefont {K.}~\bibnamefont
  {Nomura}}, \bibinfo {author} {\bibfnamefont {R.}~\bibnamefont
  {Rodr\'{\i}guez-Guzm\'an}}, \ and\ \bibinfo {author} {\bibfnamefont {L.~M.}\
  \bibnamefont {Robledo}},\ }\href@noop {} {\bibfield  {journal} {\bibinfo
  {journal} {Phys. Rev. C}\ }\textbf {\bibinfo {volume} {87}},\ \bibinfo
  {pages} {064313} (\bibinfo {year} {2013})}\BibitemShut {NoStop}%
\bibitem [{\citenamefont {Nomura}\ \emph {et~al.}(2010)\citenamefont {Nomura},
  \citenamefont {Shimizu},\ and\ \citenamefont {Otsuka}}]{nomura2010}%
  \BibitemOpen
  \bibfield  {author} {\bibinfo {author} {\bibfnamefont {K.}~\bibnamefont
  {Nomura}}, \bibinfo {author} {\bibfnamefont {N.}~\bibnamefont {Shimizu}}, \
  and\ \bibinfo {author} {\bibfnamefont {T.}~\bibnamefont {Otsuka}},\ }\href
  {\doibase 10.1103/PhysRevC.81.044307} {\bibfield  {journal} {\bibinfo
  {journal} {Phys. Rev. C}\ }\textbf {\bibinfo {volume} {81}},\ \bibinfo
  {pages} {044307} (\bibinfo {year} {2010})}\BibitemShut {NoStop}%
\bibitem [{\citenamefont {Bengtsson}\ \emph {et~al.}(1987)\citenamefont
  {Bengtsson}, \citenamefont {Bengtsson}, \citenamefont {Dudek}, \citenamefont
  {Leander}, \citenamefont {Nazarewicz},\ and\ \citenamefont
  {ye~Zhang}}]{bengtsson1987}%
  \BibitemOpen
  \bibfield  {author} {\bibinfo {author} {\bibfnamefont {R.}~\bibnamefont
  {Bengtsson}}, \bibinfo {author} {\bibfnamefont {T.}~\bibnamefont
  {Bengtsson}}, \bibinfo {author} {\bibfnamefont {J.}~\bibnamefont {Dudek}},
  \bibinfo {author} {\bibfnamefont {G.}~\bibnamefont {Leander}}, \bibinfo
  {author} {\bibfnamefont {W.}~\bibnamefont {Nazarewicz}}, \ and\ \bibinfo
  {author} {\bibfnamefont {J.}~\bibnamefont {ye~Zhang}},\ }\href {\doibase
  http://dx.doi.org/10.1016/0370-2693(87)91406-7} {\bibfield  {journal}
  {\bibinfo  {journal} {Physics Letters B}\ }\textbf {\bibinfo {volume}
  {183}},\ \bibinfo {pages} {1 } (\bibinfo {year} {1987})}\BibitemShut
  {NoStop}%
\bibitem [{\citenamefont {Bengtsson}\ and\ \citenamefont
  {Nazarewicz}(1989)}]{bengtsson1989}%
  \BibitemOpen
  \bibfield  {author} {\bibinfo {author} {\bibfnamefont {R.}~\bibnamefont
  {Bengtsson}}\ and\ \bibinfo {author} {\bibfnamefont {W.}~\bibnamefont
  {Nazarewicz}},\ }\href {\doibase 10.1007/BF01284554} {\bibfield  {journal}
  {\bibinfo  {journal} {Z. Phys. A}\ }\textbf {\bibinfo {volume} {334}},\
  \bibinfo {pages} {269} (\bibinfo {year} {1989})}\BibitemShut {NoStop}%
\bibitem [{\citenamefont {Nazarewicz}(1993)}]{nazarewicz1993}%
  \BibitemOpen
  \bibfield  {author} {\bibinfo {author} {\bibfnamefont {W.}~\bibnamefont
  {Nazarewicz}},\ }\href@noop {} {\bibfield  {journal} {\bibinfo  {journal}
  {Phys. Lett. B}\ }\textbf {\bibinfo {volume} {305}},\ \bibinfo {pages} {195 }
  (\bibinfo {year} {1993})}\BibitemShut {NoStop}%
\bibitem [{\citenamefont {{P. Van Isacker}}()}]{IBM1}%
  \BibitemOpen
  \bibfield  {author} {\bibinfo {author} {\bibnamefont {{P. Van Isacker}}},\
  }\href@noop {} {}\bibinfo {note} {Computer program IBM-1
  (unpublished)}\BibitemShut {NoStop}%
\bibitem [{\citenamefont {{Brookhaven National Nuclear Data Center}}()}]{data}%
  \BibitemOpen
  \bibfield  {author} {\bibinfo {author} {\bibnamefont {{Brookhaven National
  Nuclear Data Center}}},\ }\href@noop {} {}\bibinfo {howpublished}
  {{http://www.nndc.bnl.gov}}\BibitemShut {NoStop}%
\bibitem [{\citenamefont {Kib\'edi}\ and\ \citenamefont
  {Spear}(2005)}]{kibedi2005}%
  \BibitemOpen
  \bibfield  {author} {\bibinfo {author} {\bibfnamefont {T.}~\bibnamefont
  {Kib\'edi}}\ and\ \bibinfo {author} {\bibfnamefont {R.}~\bibnamefont
  {Spear}},\ }\href {\doibase 10.1016/j.adt.2004.11.002} {\bibfield  {journal}
  {\bibinfo  {journal} {At. Data and Nucl. Data Tables}\ }\textbf {\bibinfo
  {volume} {89}},\ \bibinfo {pages} {77 } (\bibinfo {year} {2005})}\BibitemShut
  {NoStop}%
\bibitem [{\citenamefont {Ljungvall}\ \emph {et~al.}(2008)\citenamefont
  {Ljungvall}, \citenamefont {G\"orgen}, \citenamefont {Girod}, \citenamefont
  {Delaroche}, \citenamefont {Dewald}, \citenamefont {Dossat}, \citenamefont
  {Farnea}, \citenamefont {Korten}, \citenamefont {Melon}, \citenamefont
  {Menegazzo}, \citenamefont {Obertelli}, \citenamefont {Orlandi},
  \citenamefont {Petkov}, \citenamefont {Pissulla}, \citenamefont {Siem},
  \citenamefont {Singh}, \citenamefont {Srebrny}, \citenamefont {Theisen},
  \citenamefont {Ur}, \citenamefont {Valiente-Dob\'on}, \citenamefont {Zell},\
  and\ \citenamefont {Zieli\ifmmode~\acute{n}\else
  \'{n}\fi{}ska}}]{PhysRevLett.100.102502}%
  \BibitemOpen
  \bibfield  {author} {\bibinfo {author} {\bibfnamefont {J.}~\bibnamefont
  {Ljungvall}}, \bibinfo {author} {\bibfnamefont {A.}~\bibnamefont {G\"orgen}},
  \bibinfo {author} {\bibfnamefont {M.}~\bibnamefont {Girod}}, \bibinfo
  {author} {\bibfnamefont {J.-P.}\ \bibnamefont {Delaroche}}, \bibinfo {author}
  {\bibfnamefont {A.}~\bibnamefont {Dewald}}, \bibinfo {author} {\bibfnamefont
  {C.}~\bibnamefont {Dossat}}, \bibinfo {author} {\bibfnamefont
  {E.}~\bibnamefont {Farnea}}, \bibinfo {author} {\bibfnamefont
  {W.}~\bibnamefont {Korten}}, \bibinfo {author} {\bibfnamefont
  {B.}~\bibnamefont {Melon}}, \bibinfo {author} {\bibfnamefont
  {R.}~\bibnamefont {Menegazzo}}, \bibinfo {author} {\bibfnamefont
  {A.}~\bibnamefont {Obertelli}}, \bibinfo {author} {\bibfnamefont
  {R.}~\bibnamefont {Orlandi}}, \bibinfo {author} {\bibfnamefont
  {P.}~\bibnamefont {Petkov}}, \bibinfo {author} {\bibfnamefont
  {T.}~\bibnamefont {Pissulla}}, \bibinfo {author} {\bibfnamefont
  {S.}~\bibnamefont {Siem}}, \bibinfo {author} {\bibfnamefont {R.~P.}\
  \bibnamefont {Singh}}, \bibinfo {author} {\bibfnamefont {J.}~\bibnamefont
  {Srebrny}}, \bibinfo {author} {\bibfnamefont {C.}~\bibnamefont {Theisen}},
  \bibinfo {author} {\bibfnamefont {C.~A.}\ \bibnamefont {Ur}}, \bibinfo
  {author} {\bibfnamefont {J.~J.}\ \bibnamefont {Valiente-Dob\'on}}, \bibinfo
  {author} {\bibfnamefont {K.~O.}\ \bibnamefont {Zell}}, \ and\ \bibinfo
  {author} {\bibfnamefont {M.}~\bibnamefont {Zieli\ifmmode~\acute{n}\else
  \'{n}\fi{}ska}},\ }\href {\doibase 10.1103/PhysRevLett.100.102502} {\bibfield
   {journal} {\bibinfo  {journal} {Phys. Rev. Lett.}\ }\textbf {\bibinfo
  {volume} {100}},\ \bibinfo {pages} {102502} (\bibinfo {year}
  {2008})}\BibitemShut {NoStop}%
\bibitem [{\citenamefont {Stone}(2005)}]{stone2005}%
  \BibitemOpen
  \bibfield  {author} {\bibinfo {author} {\bibfnamefont {N.}~\bibnamefont
  {Stone}},\ }\href@noop {} {\bibfield  {journal} {\bibinfo  {journal} {At.
  Data Nucl. Data Tables}\ }\textbf {\bibinfo {volume} {90}},\ \bibinfo {pages}
  {75} (\bibinfo {year} {2005})}\BibitemShut {NoStop}%
\bibitem [{\citenamefont {Wilets}\ and\ \citenamefont
  {Jean}(1956)}]{wilets1956}%
  \BibitemOpen
  \bibfield  {author} {\bibinfo {author} {\bibfnamefont {L.}~\bibnamefont
  {Wilets}}\ and\ \bibinfo {author} {\bibfnamefont {M.}~\bibnamefont {Jean}},\
  }\href@noop {} {\bibfield  {journal} {\bibinfo  {journal} {Phys. Rev.}\
  }\textbf {\bibinfo {volume} {102}},\ \bibinfo {pages} {788} (\bibinfo {year}
  {1956})}\BibitemShut {NoStop}%
\bibitem [{\citenamefont {Davydov}\ and\ \citenamefont
  {Filippov}(1958)}]{davydov1958}%
  \BibitemOpen
  \bibfield  {author} {\bibinfo {author} {\bibfnamefont {A.~S.}\ \bibnamefont
  {Davydov}}\ and\ \bibinfo {author} {\bibfnamefont {G.~F.}\ \bibnamefont
  {Filippov}},\ }\href {\doibase 10.1016/0029-5582(58)90153-6} {\bibfield
  {journal} {\bibinfo  {journal} {Nucl. Phys.}\ }\textbf {\bibinfo {volume}
  {8}},\ \bibinfo {pages} {237 } (\bibinfo {year} {1958})}\BibitemShut
  {NoStop}%
\bibitem [{\citenamefont {Elliot}\ and\ \citenamefont
  {White}(1980)}]{elliot1980}%
  \BibitemOpen
  \bibfield  {author} {\bibinfo {author} {\bibfnamefont {J.}~\bibnamefont
  {Elliot}}\ and\ \bibinfo {author} {\bibfnamefont {A.}~\bibnamefont {White}},\
  }\href {\doibase http://dx.doi.org/10.1016/0370-2693(80)90573-0} {\bibfield
  {journal} {\bibinfo  {journal} {Physics Letters B}\ }\textbf {\bibinfo
  {volume} {97}},\ \bibinfo {pages} {169 } (\bibinfo {year}
  {1980})}\BibitemShut {NoStop}%
\bibitem [{\citenamefont {Elliot}\ and\ \citenamefont
  {Evans}(1981)}]{elliot1981}%
  \BibitemOpen
  \bibfield  {author} {\bibinfo {author} {\bibfnamefont {J.}~\bibnamefont
  {Elliot}}\ and\ \bibinfo {author} {\bibfnamefont {J.}~\bibnamefont {Evans}},\
  }\href {\doibase http://dx.doi.org/10.1016/0370-2693(81)90297-5} {\bibfield
  {journal} {\bibinfo  {journal} {Physics Letters B}\ }\textbf {\bibinfo
  {volume} {101}},\ \bibinfo {pages} {216 } (\bibinfo {year}
  {1981})}\BibitemShut {NoStop}%
\end{thebibliography}%

\end{document}